\documentclass[twocolumn]{autart}    

\usepackage{cite}
\usepackage{graphicx,subfigure}
\usepackage{epsfig} 
\usepackage{amsmath} 
\usepackage{amssymb}  
\usepackage{euscript}
\usepackage{color}
\usepackage{ccaption}
\usepackage{listings}
\usepackage{booktabs} 
\usepackage{longtable}
\usepackage{multirow}
\usepackage{graphicx}
\usepackage{epstopdf}
\usepackage{bm}
\usepackage{makecell}
\usepackage{subfig}
\usepackage{algorithm}
\usepackage{algorithmicx}
\usepackage{algpseudocode}
\usepackage{tikz}
\usetikzlibrary{shapes,arrows}
\usepackage{wrapfig}

\newtheorem{theorem}{\bf \text{Theorem}}
\newtheorem{definition}{\bf \text{Definition}}

\newtheorem{assumption}{\bf \text{Assumption}}

\newtheorem{lemma}{\bf \text{Lemma}}
\newtheorem{remark}{\bf \text{Remark}}

\newcommand{\N}{{\mathbb N}}

\newcommand{\TR}{\text{R}}

\newcommand{\Trace}{\text{Tr}}

\DeclareMathOperator*{\subject}{subj. to\ }

\DeclareMathOperator*{\bdiag}{blkdiag}

\DeclareMathOperator*\argmin{arg\,min}
\DeclareMathOperator*\argmax{arg\,max}

\def\pf{\textit{Proof. }}
\graphicspath{{./results_fig/}}

\begin{document}
	
	\begin{frontmatter}
		
		\title{Kernel-based Regularized Iterative Learning Control of Repetitive Linear Time-varying Systems \thanksref{footnoteinfo}} 
		
		\thanks[footnoteinfo]{A preliminary version of this work \cite{YCML2021} was published in the 19th IFAC Symposium on System Identification (SYSID), 2021. Corresponding author: Tianshi Chen.\\ This work is supported by the youth project funded by NSFC under contract 62203375, the Thousand Youth Talents Plan funded by the central government of China, the general project funded by NSFC under contract No. 61773329, the Shenzhen research projects funded by the Shenzhen Science and Technology Innovation Council under contract No. Ji-20170189, the President's grant under contract No. PF. 01.000249 and the Start-up grant under contract No. 2014.0003.23 funded by CUHKSZ, and the Strategic Priority Research Program under Grant No. XDA27000000 funded by the Chinese Academy of Sciences.}
		
		\author[CUHKSZ]{Xian Yu}\ead{yuxian@cuhk.edu.cn},    
	\author[CUHKSZ]{Xiaozhu Fang}\ead{xiaozhufang@link.cuhk.edu.cn},
		\author[AMSS]{Biqiang Mu}\ead{bqmu@amss.ac.cn},               
		\author[CUHKSZ]{Tianshi Chen}\ead{tschen@cuhk.edu.cn}               
		
		\address[CUHKSZ]{School of Data Science and Shenzhen Research Institute of Big Data, The Chinese University of Hong Kong, Shenzhen 518172, China}
		\address[AMSS]{Key Laboratory of Systems and Control, Institute of Systems Science, Academy of Mathematics and Systems Science, Chinese Academy of Sciences, Beijing~100190, China}

		
		\vspace*{-2mm}
		\begin{abstract}                          
	   For data-driven iterative learning control (ILC) methods, both the model estimation and controller design problems are converted to parameter estimation problems for some chosen model structures. It is well-known that if the model order is not chosen carefully, models with either large variance or large bias would be resulted, which is one of the obstacles to further improve the modeling and tracking performances of data-driven ILC in practice. An emerging trend in the system identification community to deal with this issue is using regularization instead of the statistical tests, e.g., AIC, BIC, and one of the representatives is the so-called kernel-based regularization method (KRM). In this paper, we integrate KRM into data-driven ILC to handle a class of repetitive linear time-varying systems, and moreover, we show that the proposed method has ultimately bounded tracking error in the iteration domain. The numerical simulation results show that in contrast with the least squares method and some existing data-driven ILC methods, the proposed one can give faster convergence speed, better accuracy and robustness in terms of the tracking performance. 
		\end{abstract}
		
	\end{frontmatter}
	
\section{Introduction}
Iterative learning control (ILC) is a valuable control strategy for repetitive systems that operate from iteration to iteration. In practice, many systems or processes can be seen as repetitive systems, such as the batch processes in chemical industries, e.g., \cite{CHJH2018}, robotic manipulators in manufacturing lines, e.g., \cite {ZLXF2014}, and high-speed trains operated from one station to another station, e.g., \cite{YHX2018}. The learning capability from previous iterations turns the control problems, e.g., high tracking performance, of repetitive time-varying systems into a reality, and numerous ILC methods have been proposed, such as the contraction mapping based ILC, e.g., \cite{MM2020}, composite energy function based ILC, e.g., \cite{J2018}, and data-driven ILC, e.g., \cite{dRO2019, HCG2016, YuX22TC}. For systems that do not operate in repetitive manner, there are also many learning-based and data-based control methods, e.g., \cite{WZ2021, BCCGB2022}.

This paper focuses on data-driven ILC methods. The first step for most of these methods, e.g., \cite{HCG2016}, is to estimate a parametric model by using the classical system identification method, i.e., the maximum likelihood/prediction error method (ML/PEM), e.g., \cite{Ljung1999}, and to our best knowledge, the model order is often chosen to be small for ease of applications. It is worth to note that there are many methods for the identification of time-varying systems, e.g., the frequency domain ones \cite{LPL2012, GLLP2017} and reinforcement learning ones \cite{LWZ2020}. These methods often estimate the model parameters under the framework of ML/PEM, and assume that the model order is known.  There are also many traditional ILC methods for the control of repetitive time-varying systems, e.g., \cite{JX2013, LLK2000}, which either assume that the model of underlying system is known or consider the state-space model with unknown time-varying parameters to be estimated, and assume that the model order is known. However, it is well-known from the classical system identification theory that the model order selection determines the model complexity and has to be chosen carefully. If the model order is chosen to be larger than necessary, these methods would lead to model estimates with large variance, and if otherwise, with large bias. The second step for most of these methods, e.g., \cite{dRO2019, YuX22TC}, is to design a parametric learning controller based on the model estimate obtained in the first step and the available data. However, the model order (i.e., the number of controller parameters) is typically determined in an ad-hoc way. Such treatment will lead to the issue similar to the model estimation problem, i.e., if the model order is not chosen carefully, the learning controller will lead to the tracking errors with either large variance or large bias.

In fact, ML/PEM handles the bias-variance tradeoff, i.e., the model order selection, by various statistical tests, such as AIC, BIC and cross validation methods. However, they are not as reliable as expected, especially, when the data is short or has low signal-to-noise ratio, e.g., \cite{COL2012,PDCDL2014}. In the past decade, a so-called kernel-based regularization method (KRM) has emerged as a complement to ML/PEM and new system identification paradigm \cite{LCM20}. 
The success of KRM depends on the design of a suitable kernel, which is regarding how to parameterize a kernel with a parameter called hyper-parameter and embed the prior knowledge in the designed kernel, e.g., \cite{C2018}, and also on the hyper-parameter estimation, which is regarding how to estimate the hyper-parameter based on the data, e.g., \cite{MCL2018}. Different from ML/PEM, the kernel determines the underlying model structure and the hyper-parameter determines the model complexity, which can actually be tuned in a continuous and automatic way. Extensive results have shown that KRM can give model estimates with better accuracy and robustness than ML/PEM, e.g., \cite{PDCDL2014}.

In this paper, we integrate KRM into data-driven ILC methods for repetitive linear time-varying systems, where KRM is applied for both the model estimation and controller design problems. The integration of KRM opens the door for systematic use of prior knowledge of the underlying system and controller in the date-driven ILC. In particular, instead to select suitable model orders as often did before, we first set them to be sufficiently large numbers, then we choose kernels according to the prior knowledge of the underlying system and controller, and then estimate the hyper-parameter for the model estimation by using the Stein's unbiased risk estimation (SURE) method, e.g., \cite{MCL2018}, and that for the controller design by using a method that is adapted from the SURE method, and finally, we obtain the model estimate and the learning controller by solving the regularized least squares (RLS) problems. It is worth to stress that the proposed method has all of its parameters tuned in an automatic way, and thus is different from the existing data-driven ILC methods, e.g., \cite{dRO2019, YuX22TC}, where some parameters need to be tuned in an ad-hoc way. The proposed method does not include the model order selection problem, the model order only needs to be chosen sufficiently large to capture the dynamics of underlying system and controller and the model complexity is tuned by the hyper-parameter in a continuous and automatic way, and thus is different from the methods in \cite{LPL2012, GLLP2017, LWZ2020, JX2013, LLK2000}, which assume that the model order is known. Moreover, we prove that the tracking error for the proposed method is ultimately bounded in the iteration domain. Finally, we run numerical simulations to show that the proposed method can give faster convergence speed, better accuracy and robustness in terms of the tracking performance in comparison with the least squares (LS) method, and the data-driven ILC methods in \cite{dRO2019, YuX22TC}.

The remaining part of this paper is organized as follows. In Section \ref{sec2}, we give the problem statement and some preliminary results. In Section \ref{sec3}, we show how to integrate KRM into data-driven ILC. In Section \ref{sec4}, we show that the tracking error of the proposed method is ultimately bounded, and then in Section \ref{sec5}, we show the numerical simulation, and finally, we conclude this paper in Section \ref{sec6}.



\section{Problem Statement and Preliminaries}\label{sec2}
In this paper, we consider repetitive discrete-time linear time-varying systems described by the following linear time-varying ARX model:
\begin{subequations}\label{equ:system}
  \begin{align}
  & A(q,t) y_j(t) = B(q,t) u_j(t) + v_j(t), \\
  & A(q,t) = 1 + \sum\limits_{l=1}^{n_a} a_l(t) q^{-l}, \\
  & B(q,t) = \sum\limits_{k=1}^{n_b} b_k(t) q^{-k},\\
  & t=1,\cdots,N_d,\quad j=1,\cdots,N_e,
  \end{align}
\end{subequations}
where $t \in \mathbb{N}$ is the time instant, $j \in \mathbb{N}$ is the iteration (experiment) index at which the model (\ref{equ:system}) repeats, $q$ is the time shift operator such that $qy_j(t) = y_j(t+1)$, $y_j(t),u_j(t),v_j(t) \in \mathbb{R}$ are the output, input and measurement noise at iteration $j$ and time $t$, respectively, and moreover $v_j(t)$ is i.i.d. (with respect to both $j$ and $t$) with zero mean, variance $\sigma^2$ and constraint $| v_j(t) | \leq d_v$,  $d_v >0$ is a constant, $a_l (t) \in \mathbb{R}$, $l=1, \cdots, n_a$, and $b_k(t) \in \mathbb{R}$, $k=1, \cdots, n_b$, are the unknown time-varying parameters of the model \eqref{equ:system}, $n_a, n_b \in \mathbb{N}$ are the orders of the model \eqref{equ:system}, and $N_d, N_e\in\mathbb N$ are the number of data and iterations, respectively.

The time-varying system (\ref{equ:system}) can be rewritten in the following linear regression form:
\begin{subequations}\label{equ:system-regression}
\begin{align}
& y_j(t) =\varphi_j^T(t) \theta_m(t)+ v_j(t), \\
& \theta_m(t) = \left[\theta_b^T(t), \theta_a^T(t)  \right]^T,  \\
&  \theta_b(t) = [b_1(t), \cdots, b_{n_b}(t)]^T,  \\
&  \theta_a(t) = [a_1(t), \cdots, a_{n_a}(t)]^T,\\
& \varphi_{j}(t) =  \big[ u_j(t-1), \cdots, u_j(t-n_b), \nonumber \\
& \quad \quad\quad\;\; -y_j(t-1), \cdots, -y_j(t-n_a)  \big]^T,\\
& t=1,\cdots,N_d, \quad j=1,\cdots,N_e,
\end{align}
\end{subequations}
where $\theta_m(t) \in \mathbb{R}^{n_a+n_b}$ is the unknown model parameter to be estimated at time $t$, which contains two parts: the parameter of $B(q,t)$ denoted by $\theta_b(t)$,  and the parameter of $A(q,t)$, denoted by $\theta_a(t)$, and $\varphi_{j}(t) \in \mathbb{R}^{n_a+n_b}$ is the regressor at iteration $j$  and time $t$.


\subsection{Problem Statement}\label{subsec2-1}
For each iteration $j=1,\cdots,N_e$ and time $t=1,\cdots,N_d$, the data-driven ILC for the time-varying system (\ref{equ:system}) contains the following two tasks:
\begin{description}
  \item[1) model estimation:] estimate the model parameter $\theta_m(t)$ (the corresponding estimate is denoted by $\hat{\theta}_{m,j}(t)$) as well as possible based on the data $\{u_{i}(t-1), \cdots, u_{i}(t-n_b), y_{i}(t), \cdots, y_{i}(t-n_a) \}_{i=1}^{j-1}$.

\vspace{2mm}
  \item[2) controller design:] for the reference output $y_d(t+1)$, find an input $u_j(t)$ based on the model estimate $\hat{\theta}_{m,j}(t)$ and the data $\{u_{i}(t-1), \cdots, u_{i}(t-n_b), y_{i}(t), \cdots, y_{i}(t-n_a) \}_{i=1}^{j-1}$ and $\{u_j(k-1), \cdots, u_j(\\k - n_b), y_j(k), \cdots, y_j (k - n_a) \}_{k=1}^{t}$, such that the tracking error
\begin{align}\label{equ:error}
 e_j(t+1) = y_d(t+1)-y_j(t+1),
\end{align}
is as small as possible, where $y_j(0)= y_j(1)=0$ and $u_j(0)=0$\footnote{It should be noted that for ILC, it is common to assume that the initial conditions $y_j(0), y_j(1), u_j(0)$, and the reference output $y_d(t+1)$ remain unchanged for each iteration, e.g., \cite{dRO2019}. }.
\end{description}




For both the model estimation and controller design,  data-driven ILC methods, e.g., \cite{HCG2016,YuX22TC}, often estimate a parametric model by using ML/PEM, where the model orders $n_a$ and $n_b$ of system (\ref{equ:system}), and $n_c$ in the learning controller (\ref{equ:ILC-law}) are chosen in an ad-hoc way. However, it is well-known that if these numbers are chosen to be larger than necessary, ML/PEM would lead to model/controller estimates with large variance, and if otherwise, with large bias. Actually, in the classical system identification, these parameters are often chosen by using various statistical tests, such as AIC, BIC and cross validation methods, e.g., \cite{Ljung1999}, but they are not as reliable as expected, especially when the data is short or has low signal-to-noise ratio, e.g., \cite{COL2012, PDCDL2014}. In the last decade, the so-called KRM has emerged as a complement to ML/PEM and new system identification paradigm \cite{LCM20}, and one of its novelties is that it provides a different route to handle the bias-variance tradeoff by using the regularization, which embeds the prior knowledge of underlying system.

\subsection{Kernel-based Regularized Least Squares Method}\label{subsec2-2}

In this section, we briefly review the kernel-based regularization method (KRM).
We consider the following linear regression model:
\begin{equation}\label{equ:model-eg}
\begin{aligned}
  Y =\Phi \theta + V,
\end{aligned}
\end{equation}
where $Y, V \in \mathbb{R}^N$, $\Phi \in \mathbb{R}^{N \times n}$, and $\theta \in \mathbb{R}^n$ are the measurement output, measurement noise, regression matrix and model parameter, respectively.

The KRM estimates $\theta$ by minimizing the following regularized least squares (RLS) criterion:
\begin{subequations}\label{equ:estimator-RLS-eg}
\begin{align}
  &  \hat{\theta}^{\TR} =  \argmin \limits_{\theta \in \mathbb{R}^{n}} \| Y - \Phi \theta \|^2 + \sigma^2 \theta^T P^{-1} \theta, \\
  &  \quad\,  = \left( \Phi^T \Phi + \sigma^2 P^{-1}  \right)^{-1} \Phi^T Y,
\end{align}
\end{subequations}
where $\|\cdot\|$ represents the Euclidean norm of a vector or the spectral norm of a matrix, $P \in  \mathbb{R}^{n \times n}$ is a positive semidefinite matrix, e.g., \cite{PDCDL2014}. Obviously, the RLS estimator $\hat{\theta}^{\TR}$ in (\ref{equ:estimator-RLS-eg}) becomes the least squares (LS) estimator when $P^{-1} = 0$.

The matrix $P$ determines the underlying model structure of $\hat{\theta}^{\TR}$, e.g., \cite{PDCDL2014}, thus has to be designed carefully and is parameterized by the so-called hyper-parameter $\eta \in \Omega \subset \mathbb{R}^{n_\eta}$ with $n_{\eta} \in \mathbb{N}$, and in this case, we rewrite $P$ as $P(\eta)$. The most popular way to design $P(\eta)$ is through a positive semidefinite kernel and in this case, $P(\eta)$ is often called the kernel matrix. For example, if $\theta$ represents a vector of finite impulse response coefficients, the $(k, l)$th element of $P(\eta)$, i.e., $P_{kl}(\eta)$ can be chosen as the diagonal correlated (DC) kernel
\begin{subequations}\label{equ:kernel}
\begin{align}
  & \text{DC}: P_{kl}(\eta) = c \alpha^{(k+l)/2} \beta^{| k-l |}, \nonumber \\
  & \quad\quad\; \eta= [c, \alpha, \beta] \in \Omega  = \{ c \geq 0, 0 \leq \alpha < 1, |\beta| \leq 1 \},  
\end{align} where $c,\alpha$ and $\beta$ describe the scaling factor, decay rate and correlation of $\theta$, respectively. The DC kernel and its special cases are widely used in the regularized system identification, e.g., the so-called tuned-correlated (TC) kernel and the diagonal (DI) kernel \cite{COL2012}
\begin{align}
  & \text{TC}: P_{kl}(\eta) = c \alpha^{\max(k,l)},  \nonumber \\
  & \quad\quad\; \eta= [c, \alpha] \in \Omega  = \{ c \geq 0, 0 \leq \alpha < 1 \},\\
  & \text{DI}: P_{kl}(\eta) =\left\{\begin{array}{cc}
  c \alpha^k, & k=l, \\
  0, & k\neq l,
  \end{array} \right.   \nonumber \\
  & \quad\quad\; \eta= [c, \alpha] \in \Omega  = \{ c \geq 0, 0 \leq \alpha < 1 \},
\end{align}
\end{subequations}
 which are obtained from (6a) with $\beta = \alpha^{\frac{1}{2}}$ and $\beta=0$, respectively.
Once the kernel matrix $P(\eta)$ is fixed, the next task is to estimate the hyper-parameter $\eta$ based on the data. Since the hyper-parameter determines the model complexity, the hyper-parameter estimation in essence handles the bias-variance tradeoff, e.g., \cite{MCL2018}. There are many methods to complete this task, e.g., the empirical Bayes method, the Stein's unbiased risk estimation (SURE) method, and the generalized cross validation method, e.g., \cite{MCL2018, PDCDL2014}. Here, we choose to use the SURE method, because it does not require the Gaussian assumption on the measurement noise and moreover, it can keep in line with the case, where $\theta$ is subject to some quadratic constraints as required in the controller design in Section \ref{subsec3-2}.
The SURE method for the RLS estimator $\hat{\theta}^{\TR}$ in \eqref{equ:estimator-RLS-eg} takes the form of
\begin{align}\label{equ:empirical-Bayes-eg}
   &  \hat{\eta} = \argmin \limits_{\eta \in \Omega}   || Y - \Phi \hat{\theta}^{\TR}(\eta)  ||^2\nonumber\\
   & \;\;\,+ 2 \sigma^2  \Trace  \left( \Phi  \left(\Phi^T \Phi+ \sigma^2  P^{-1} (\eta) \right)^{-1} \Phi^T\right),
\end{align}
where $\Trace(\cdot)$ is the trace of a square matrix.


%
%
%

\subsection{Uniform Bounded-input Bounded-output Stability of Linear Time-varying State-space Models}\label{subsec2-3}
We consider discrete-time linear time-varying systems described by the following linear state-space model:
\begin{subequations}\label{equ:state-model-eg}
\begin{align}
  & x(t+1) = \mathcal{A}(t) x(t) +\mathcal{B}(t) u(t), \\
  & y (t) = \mathcal{C} (t) x(t),
\end{align}
\end{subequations}
where $x(t) \in \mathbb{R}^{n_x}$, $u(t) \in \mathbb{R}^{n_u}$, $y (t) \in \mathbb{R}^{n_y}$ are the state, input and output, respectively, with $n_x, n_u, n_y \in \mathbb{N}$,  and $ \mathcal{A}(t) \in \mathbb{R}^{n_x \times n_x}$, $\mathcal{B}(t) \in \mathbb{R}^{n_x \times n_u}$, $\mathcal{C} \in \mathbb{R}^{n_y \times n_x}$ are the system, input and output matrices, respectively.

\begin{definition}\cite[Definition 27.1]{R1996}\label{def:BIBO-stability-eg}
The linear state-space model \eqref{equ:state-model-eg} is said to be uniformly bounded-input, bounded-output (BIBO) stable if there exists a finite constant $d_g$ such that for any input $u(t)$, the zero-state response of \eqref{equ:state-model-eg} satisfies
 \begin{align}\label{equ:bounded-IO-eg}
   \sup\limits_{t \geq 1} \| y(t) \| \leq d_g \sup\limits_{t \geq 1} \| u(t) \|.
\end{align}
\end{definition}

A necessary and sufficient condition for systems in the form of  \eqref{equ:state-model-eg} to be uniformly BIBO stable is given by the following lemma.

\begin{lemma}\cite[Theorem 27.2]{R1996}\label{lmm:BIBO-response-eg} Consider the linear state-space model \eqref{equ:state-model-eg}. Then its zero-state response satisfies
\begin{align}
y(t)=\sum_{i=0}^{t-1}G(t,i)u(i),
\end{align} where $G(t,i)$ for all $t,i \in \mathbb{N}$ such that $t\geq i+1$ is the impulse response of \eqref{equ:state-model-eg} and takes the following form:
\begin{align}\label{eq:impulse response of linear time varying system}
   & G(t,i) = \mathcal{C}(t) \Psi(t,i+1) \mathcal{B}(i), \nonumber \\
   & \Psi(t,i) = \left \{\begin{aligned}
      & \mathcal{A}(t-1)\mathcal{A}(t-2) \cdots \mathcal{A}(i),  & t \geq i+1,  \\
      &  I_{n_x},  & t=i,
     \end{aligned} \right.
\end{align} where $I_{n_x}$ denotes the $n_x$-dimensional identity matrix. Moreover, the linear state-space model \eqref{equ:state-model-eg} is uniformly BIBO stable if and only if there exists a finite constant $d_g$ such that the impulse response $G(t,i)$ satisfies
\label{equ:bounded-G-eg}
 \begin{align}
   \sum\limits_{i=0}^{t-1} \| G(t,i) \| \leq d_g,
\end{align}
for all $t,i \in \mathbb{N}$ with $t \geq i+1$.

\end{lemma}


\section{Kernel-based Regularized ILC}\label{sec3}
In this section, we show how to integrate KRM into ILC. In particular, we first consider the model estimation, and then the controller design, and finally, we summarize the proposed kernel-based regularized ILC method.

\subsection{Model Estimation}\label{subsec3-1}

For each iteration $j=1,\cdots,N_e$ and time $t=1,\cdots,N_d$, the RLS estimate $\hat{\theta}_{m,j}^{\TR} (t)$ of $\theta_m(t)$ can be obtained by
\begin{subequations}\label{equ:estimator-RLS-model}
\begin{align}
  & \hat{ \theta}_{m,j}^{\TR} (t) = \argmin \limits_{\theta_m(t) \in \mathbb{R}^{n_a+n_b}} \left\| Y_{j-1}(t) - \Phi_{j-1}(t) \theta_m(t) \right\|^2  \nonumber \\
  & \quad\quad\quad+ \sigma^2 \theta_m^T(t) P_m^{-1} (t) \theta_m(t),  \\
 & = \left( \Phi_{j-1}^T(t) \Phi_{j-1}(t) + \sigma^2 P_m^{-1}(t) \right)^{-1} \Phi_{j-1}^T(t) Y_{j-1}(t),
\end{align}
\end{subequations}
where  $Y_{j-1}(t) = [ y_1(t), \cdots, y_{j-1}(t) ]^T$, $\Phi_{j-1}(t) = [ \varphi_{1}(t), \cdots, \varphi_{j-1}(t)  ]^T$, and $P_m(t) \in \mathbb{R}^{(n_a+n_b) \times (n_a+n_b)}$ is the time-varying kernel matrix of $\theta_m(t)$.

\begin{remark}
The model estimation \eqref{equ:estimator-RLS-model} could be handled by using the least squares method instead. However, there are three problems that should be considered in this case:
\begin{itemize}
  \item The first one is what is the number of iterations we should choose, since the number of iterations has to be larger than $n_a+n_b$.
  \item The second one is that the model parameter can not be computed at the first $n_a+n_b-1$ iterations since the model estimation problem is ill-conditioned, and only when the number of iteration is equal to or larger than $n_a+n_b$, it is possible to compute the model parameter, but we still can not guarantee that the model estimation problem is well-conditioned.
  \item The third one is how to determine the values of $n_a$ and $n_b$ since they are unknown.
\end{itemize}
\end{remark}

It should be noted that we are dealing with the identification problem of \emph{repetitive} time-varying ARX model \eqref{equ:system} and the \emph{repetitive} nature makes it different from the ordinary identification problem of linear parameter-varying (LPV) systems. For the ordinary identification problem of LPV systems, e.g., \eqref{equ:system} without the \emph{repetitive} assumption, it is standard to estimate the model parameters $[\theta_m^T(1), \cdots, \theta_m^T(N_d)]^T$ in one shot. But for the identification problem of \emph{repetitive} time-varying ARX model \eqref{equ:system}, it is also possible to estimate $\theta_m(t)$ at different time instants $t=1,\cdots,N_d$, respectively, because the initial conditions remain unchanged for each iteration and the data is collected in a \emph{repetitive} way. Moreover, the idea to estimate $\theta_m(t)$ at different time instants $t=1,\cdots,N_d$, respectively, is straightforward, and has been shown clearly in \eqref{equ:estimator-RLS-model}.

Now we consider the design of the time-varying kernel matrix $P_m(t)$. It can be concluded by the same reasoning in \cite[Section 5.3]{PDCDL2014} that for each iteration $j=1,\cdots,N_e$ and time $t=1,\cdots,N_d$, the linear time-varying ARX model (\ref{equ:system}) can be rewritten as follows
\begin{subequations}\label{equ:ARX-model-separate}
\begin{align}
& y_j(t) =\varphi_{u,j}^T(t) \theta_b(t) +\varphi_{y,j}^T(t) \theta_a(t) + v_j(t), \\
& \varphi_{u,j} (t) = [ u_j(t-1), \cdots, u_j(t-n_b) ]^T,  \\
& \varphi_{y,j} (t) = [ -y_j(t-1), \cdots, -y_j(t-n_a) ]^T,
\end{align}
\end{subequations}
where $\theta_b(t)$ and $\theta_a(t)$ are the vectors of model parameters at time $t$ associated with the input $u_j(t)$ and output $y_j(t)$, respectively, and $n_a$ and $n_b$ are often chosen in an ad hoc way, e.g., \cite{LPL2012, GLLP2017, JX2013}.

\vspace{-2ex}
Then following the suggestion in \cite{PDCDL2014}, for each time $t=1, \cdots, N_d$, we design the time-varying kernel matrix $P_m(t)$ as a block diagonal matrix with two blocks corresponding to $\theta_b(t)$ and $\theta_a(t)$, respectively. More specifically, we let $\eta_m(t)$ be the hyper-parameter used to parameterize $P_m(t)$ and then rewrite $P_m(t)$ as $P_m( \eta_m(t) )$, which takes the form of
\begin{align}
   P_m( \eta_m(t) ) = \left( \begin{array}{cc}  P_b(\eta_b(t)) & 0 \\  0 & P_a(\eta_a(t))  \end{array} \right), \nonumber
\end{align}
where $\eta_m(t) = [ \eta_b(t)^T, \eta_a(t)^T ]^T$, and $\eta_b(t)$ and $\eta_a(t)$ are the hyper-parameters used to parameterize the kernel matrices $P_b(\eta_b(t))$ of $\theta_b(t)$  and $P_a(\eta_a(t))$ of $\theta_a(t)$, respectively. The prior knowledge that is often used in practice is that $n_a$ and $n_b$ for $\theta_a(t)$ and $\theta_b(t)$ are chosen to be finite at each time $t$. Actually we can interpret the prior knowledge as follows: we let both $n_a$ and $n_b$ to be infinity, and accordingly, for any $t$, both $\theta_a(t)$ and $\theta_b(t)$ have finite $\ell_1$ norms, which implies that for any $t$, $\lim_{l\to\infty}a_l(t)=0$ and $\lim_{k\to\infty}b_k(t)=0$. In this case, $P_a(\eta_a(t))$ and $P_b(\eta_b(t))$ can be defined through the DI kernel in (\ref{equ:kernel}) since it is the simplest kernel to embed the prior knowledge, e.g., \cite{COL2012}, and moreover, for given $t$, through tuning $\alpha$, we can tune how fast $\hat{ a}_{l,j}^{\TR} (t)$ (estimate of $a_l(t)$, $l=1, \cdots, n_a$) and $\hat{ b}_{l,j}^{\TR} (t)$ (estimate of $b_l(t)$, $l=1, \cdots, n_b$) decay to zero as $l$ increases, and thus in this way, the model complexity can be tuned in a continuous and automatic way.

\vspace{-2ex}
Finally, for each iteration $j=1,\cdots,N_e$ and time $t=1, \cdots, N_d$, the estimate $\hat{\eta}_{m,j}(t)$ of $\eta_m(t)$ can be obtained by using the SURE method as follows:
\begin{align}\label{equ:hyper-parameter-estimate-model}
   & \hat{\eta}_{m,j}(t) = \argmin \limits_{\eta_m(t) \in \Omega}  \left\| Y_{j-1}(t)  - \Phi_{j-1}(t) \hat{ \theta}_{m,j}^{\TR} (t)(\eta_m(t))  \right\|^2 \nonumber \\
   & \qquad\quad + 2 \sigma^2 \Trace \Big(  \Phi_{j-1}(t) \left(\Phi_{j-1}^T(t) \Phi_{j-1}(t) \right. \nonumber \\
   & \qquad\quad \left. + \left. \sigma^2 P_m ^{-1} (\eta_m(t)) \right)^{-1}\Phi_{j-1}^T(t) \right) .
\end{align}


\vspace{-1ex}
\subsection{Controller Design}\label{subsec3-2}
\vspace{-1ex}
For each iteration $j=1,\cdots,N_e$ and time $t=1,\cdots,N_d$, we design the learning controller in the form of
\begin{subequations}\label{equ:ILC-law}
\begin{align}
  &  u_j(t) = u_{j-1}(t) + u_{c,j}(t),\\
  &  u_{c,j}(t) = E_{j-1}^T(t+1) \theta_c (t), \\
  & E_{j-1} (t+1) = [ e_{j-1} (t+1), \cdots, e_{j-n_c} (t+1) ]^T,\\
  & \theta_c (t)=[c_1(t), \cdots, c_{n_c}(t)]^T,
\end{align}
\end{subequations}
where $\theta_c (t) \in \mathbb{R}^{n_c}$ with $n_c \in \N$ is the unknown parameter of the learning controller to be estimated at time $t$, and $n_c$ is the model order (i.e., the number of controller parameters). To guarantee that the tracking error \eqref{equ:error} is ultimately bounded in the iteration domain when the designed learning controller \eqref{equ:ILC-law} is applied to the linear time-varying ARX model \eqref{equ:system} as will be studied in the next section, we assume that
for each iteration $j=1,\cdots,N_e$ and time $t=1,\cdots,N_d$,  $u_j(t)$ and $\theta_c(t)$ are bounded above by the constants $d_u>0$ and $d_c>0$, respectively, i.e., $| u_j(t) | \leq d_u$ and $\| \theta_c(t) \| \leq d_{c}$.


It is interesting to note that for each time $t=1,\cdots,N_d$, the $u_{c,j}(t)$ part in the learning controller (\ref{equ:ILC-law}) can be seen  as a linear iteration-invariant dynamic system in the iteration domain, with $e_j (t+1)$ and $ u_{c,j}(t)$ as the input and output of the system, respectively, and $\theta_c(t)$ as the finite impulse responses of the system in the iteration domain at time $t$. More specifically, this system can be shown by a block diagram in Fig. \ref{fig:controller-block},
\begin{figure}[h]
 \centering
 \resizebox{0.3\textwidth}{!}{
 \tikzstyle{block} = [draw=black, text centered, thick, rectangle, minimum height=1em, minimum width=4em]
  \tikzstyle{arrow} = [thick,->,>=stealth]
  \begin{tikzpicture}
        \node (input) [coordinate] {};
        \node (system) [block,right of =input, node distance=3cm] {$G_c(p,\theta_c(t))$};
        \node (output) [coordinate, right of =system, node distance=2.5cm] {};
        \draw [arrow] (input) -- node[anchor=south]{$e_j(t+1)$} (system);
        \draw [arrow] (system) -- node[anchor=south]{$ u_{c,j}(t)$} (output);
 \end{tikzpicture}}
 \caption{Block diagram of the learning controller (\ref{equ:ILC-law}).}\label{fig:controller-block}
\end{figure}
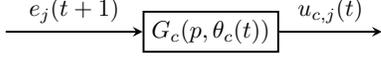\\
where
\begin{align}\label{eq:Gc}
G_c(p,\theta_c(t)) = \sum\limits_{k=1}^{n_c} c_k(t) p^{-k},
\end{align}
$p$ is the iteration shift operator such that $p^{-1} e_j(t+1) = e_{j-1}(t+1)$,and $n_c$ is often chosen in an ad hoc way, e.g., \cite{LLK2000, GN2006, NB2011}.

\vspace{-1ex}
\subsubsection{Controller Design by RLS}\label{subsec3-2-1}
\vspace{-1ex}
For the learning controller in the form of (\ref{equ:ILC-law}), the controller design problem is converted to an estimation problem of the parameter $\theta_c(t)$ of the learning controller, that is, to estimate $\theta_c(t)$ based on the data such that the tracking error \eqref{equ:error} is as small as possible.

\vspace{-1ex}
However, note that for each iteration $j=1,\cdots,N_e$ and time $t=1, \cdots, N_d$, the measurement output $y_j(t+1)$ in \eqref{equ:error} is not available, and thus its one-step-ahead predictor $\hat{y}_j(t+1)$ based on the obtained model estimate $\hat{\theta}_{m,j}^{\TR}(t+1)$ in \eqref{equ:estimator-RLS-model} is used instead, i.e.,
\begin{align}\label{equ:output-predictor-one-step}
 \hat{y}_j(t+1) = \varphi_j^T(t+1) \hat{\theta}_{m,j}^{\TR}(t+1),
\end{align}
which is further rewritten by
\begin{subequations}\label{equ:output-predictor-partioned}
\begin{align}
   & \hat{y}_j(t+1) = \bar{\varphi}_j^T(t) \hat{\theta}_{m,j}^{\TR}(t+1) + \hat{b}_{1,j}^{\TR}(t+1) u_j(t),  \\
   & \bar{\varphi}_j(t) = [0, u_j(t-1), \cdots, u_j(t-n_b+1), \nonumber \\
    &    \quad\quad\quad\; - y_j(t), \cdots, - y_j(t-n_a+1) ]^T,
\end{align}
\end{subequations}
where $\hat{b}_{1,j}^{\TR}(t+1)$ is the first element of $\hat{\theta}_{m,j}^{\TR}(t+1)$. Then substituting  \eqref{equ:ILC-law} into \eqref{equ:output-predictor-partioned}, we have
\begin{align}\label{equ:output-predictor-controller-applied}
    \hat{y}_j(t+1) & = \bar{\varphi}_j^T(t) \hat{\theta}_{m,j}^{\TR}(t+1) + \hat{b}_{1,j}^{\TR}(t+1) u_{j-1}(t) \nonumber \\
    & + \hat{b}_{1,j}^{\TR}(t+1) E_{j-1}^T (t+1) \theta_c(t).
\end{align}
Finally, with \eqref{equ:output-predictor-controller-applied}, the controller design problem is now converted to estimate $\theta_c(t)$ based on the data such that the estimated tracking error of $e_j(t+1)$ in \eqref{equ:error}, i.e.,
\begin{align}\label{equ:estimate-error}
    y_d(t+1) - \hat{y}_j(t+1),
\end{align}
is as small as possible.

\vspace{-1ex}
Accordingly, from \eqref{equ:estimate-error}, for each iteration $j=1,\cdots,N_e$ and time $t=1, \cdots, N_d$, we can obtain the following linear regression model of $\theta_c(t)$:
\begin{align}\label{equ:tracking-model-regression}
  y_{c,j} (t) =  \varphi_{c,j}^T(t) \theta_c(t) + v_{c,j}(t),
\end{align}
where\vspace{-2ex}
\begin{align}
 & y_{c,j} (t)= y_d(t+1) - \bar{\varphi}_j^T (t) \hat{\theta}_{m,j}^{\TR}(t+1) \nonumber \\
 & \qquad\;\;\, - \hat{b}_{1,j}^{\TR}(t+1) u_{j-1}(t), \nonumber \\
 & \varphi_{c,j}(t) = \hat{b}_{1,j}^{\TR}(t+1) E_{j-1} (t+1),\nonumber
\end{align}
and $v_{c,j}(t)$ is the noise and assumed to be i.i.d. (with respect to both $j$ and $t$) with zero mean and unknown variance $\sigma_c^2$.
Then for each iteration $j=1,\cdots,N_e$ and time $t=1, \cdots, N_d$, the RLS estimate $\hat{\theta}_{c,j}^{\TR}(t)$ of $\theta_c(t)$ can be obtained by\vspace{-1ex}
\begin{subequations}\label{equ:estimator-RLS-controller}
\begin{align}
  & \hat{\theta}_{c,j}^{\TR} (t) = \argmin \limits_{\theta_c \in \mathbb{R}^{n_c}} \left( y_{c,j}(t) - \varphi_{c,j}^T(t) \theta_c(t) \right)^2 \nonumber \\
  & \quad\quad\;\;\, + \sigma_c^2 \theta_c^T (t) P_c^{-1} (t) \theta_c (t), \\
  & \subject : \quad \| \theta_c (t) \| \leq d_{c}, \quad  | u_j(t) | \leq d_u,
\end{align}
\end{subequations}
where $P_c(t) \in \mathbb{R}^{n_c \times n_c}$ is the time-varying kernel matrix of $\theta_c(t)$ in (\ref{equ:ILC-law}), and the constraints are due to the ones mentioned in (\ref{equ:ILC-law}).

\begin{remark}
For the controller design \eqref{equ:estimator-RLS-controller}, if we choose to include multiple previous iterations of tracking errors which have bad tracking performances, then it is possible to get the incorrect bias-variance tradeoff, i.e., the tradeoff between the data fit and the complexity of controller. Therefore, the number of iterations of data in the cost function of the controller design \eqref{equ:estimator-RLS-controller} is not suggested to be large. If it is large, it will lead to bad tracking performance or even divergent tracking. In this paper we include one iteration of data in \eqref{equ:estimator-RLS-controller}, which is widely used in many ILC methods, e.g., \cite{J2018, JX2013}, and the simulation results in Figs. \ref{fig:control-fit-system1} and \ref{fig:control-fit} have shown that a good tracking performance is obtained.
\end{remark}

Now, the learning controller (\ref{equ:ILC-law}) takes the form of
\begin{equation}\label{equ:ILC-law-applied}
 u_j(t) = u_{j-1}(t) + E_{j-1}^T(t+1) \hat{\theta}_{c,j}^{\TR} (t).
\end{equation}
For a given time-varying kernel matrix $P_c(t)$, for each iteration $j=1,\cdots,N_e$ and time $t=1, \cdots, N_d$, the optimal solution of the constrained optimization problem \eqref{equ:estimator-RLS-controller} has to satisfy the following  \textit{Karush-Kuhn-Tucker} conditions \cite[p. 243-244]{BV2004}:
\begin{subequations}
\begin{align}
   & | E_{j-1}^T(t+1) \hat{\theta}_{c,j}^{\TR} (t) + u_{j-1}(t) |^2 - d_u^2 \leq 0, \\
   & \| \hat{\theta}_{c,j}^{\TR} (t) \|^2 -  d_{c}^2 \leq 0, \\
   & \lambda_{1,j}^*(t) \left( | E_{j-1}^T(t+1) \hat{\theta}_{c,j}^{\TR} (t) + u_{j-1}(t) |^2 - d_u^2 \right) = 0, \\
   & \lambda_{2,j}^*(t) \left( \| \hat{\theta}_{c,j}^{\TR} (t) \|^2 -  d_{c}^2 \right) = 0, \\
   & \left( \varphi_{c,j}(t) \varphi_{c,j}^T(t)  + \lambda_{1,j}^*(t) E_{j-1}(t+1) E_{j-1}^T(t+1) \right. \nonumber \\
   & \left. + \lambda_{2,j}^*(t) I_{n_c} + \sigma_c^2 P_c^{-1}(t) \right) \hat{\theta}_{c,j}^{\TR} (t) - \varphi_{c,j}(t) y_{c,j}(t)  \nonumber \\
   & + \lambda_{2,j}^*(t) u_{j-1}(t) E_{j-1}(t+1)= 0, \\
   & \lambda_{1,j}^*(t) \geq 0, \quad \lambda_{2,j}^*(t) \geq 0,
\end{align}
\end{subequations}
where $\lambda_{1,j}^*(t), \lambda_{2,j}^*(t) \geq 0$ are the dual optimal variables. Since the constrained optimization problem \eqref{equ:estimator-RLS-controller} is convex, it can be efficiently solved using, e.g., the \textit{CVX} \cite{GB2014} or the function \textit{fmincon} in Matlab.

\vspace{-1ex}
\subsubsection{Design of Kernel Matrix and Estimation of Hyper-parameters}\label{subsec3-2-2}
\vspace{-1ex}
Now we consider the design of the time-varying kernel matrix $P_c(t)$ in \eqref{equ:estimator-RLS-controller}. Similar to the model estimation case, for each time $t=1, \cdots, N_d$, the matrix $P_c(t)$ is often parameterized by a hyper-parameter $\eta_c(t) \in \Omega \subset \mathbb{R}^{n_\eta}$ with $n_{\eta} \in \mathbb{N}$ and can be rewritten as $P_c(\eta_c(t))$. The prior knowledge that is often used in practice is that $n_c$ for $\theta_c(t)$ is chosen to be finite. Actually we can interpret the prior knowledge as follows: we let $n_c$ to be infinity, and accordingly, for any $t$, $\theta_c(t)$ has finite $\ell_1$ norm, which implies that for any $t$, $\lim_{k\to\infty}c_k(t)=0$. In this case, $P_c(\eta_c(t))$ can be defined through the DI kernel in (\ref{equ:kernel}) since it is the simplest kernel to embed the prior knowledge, e.g., \cite{COL2012}, and moreover, for given $t$, through tuning $\alpha$, we can tune how fast $\hat{ c}_{l,j}^{\TR} (t)$ (estimate of $c_l(t)$, $l=1, \cdots, n_c$) decays to zero as $l$ increases, and thus in this way, the complexity of learning controller \eqref{equ:ILC-law} can be tuned in a continuous and automatic way.

\vspace{-2ex}
Then, for each time $t=1, \cdots, N_d$, we consider the estimation of the hyper-parameter $\eta_{c}(t)$. Note that the SURE method \eqref{equ:empirical-Bayes-eg} is based on the closed-form expression of the RLS estimator \eqref{equ:estimator-RLS-eg}, while the RLS estimator \eqref{equ:estimator-RLS-controller} generally has no closed-form expression due to the constraints. To tackle this difficulty,  we propose in the following a two-steps procedure to estimate $\eta_c (t)$.

\vspace{-2ex}
In the first step, following \cite[Sections 5.1 and 5.2]{BV2004}, we first minimize the Lagrangian of \eqref{equ:estimator-RLS-controller} and obtain a closed-form expression of $\theta_c(t)$ up to the dual variables, and then we maximize the Lagrange dual function of \eqref{equ:estimator-RLS-controller} to get the optimal values of the dual variables. More specifically, for the constrained optimization problem \eqref{equ:estimator-RLS-controller}, and for each iteration $j=1, \cdots, N_e$ and time $t=1, \cdots, N_d$, the Lagrangian is given by
\begin{align}\label{equ:Lagrangian-eg}
& L_j(\theta_c(t), \lambda_1(t), \lambda_2(t), \eta_c(t)) = \left( y_{c,j}(t) - \varphi_{c,j}^T(t)  \theta_{c} (t) \right)^2 \nonumber\\
&+ \sigma_c^2 \theta_{c}^T(t) P_c^{-1}(\eta_c(t)) \theta_{c}(t) \nonumber \\
&+ \lambda_1(t) \left( | E_{j-1}^T(t+1) \theta_{c}(t) + u_{j-1}(t) |^2-d_u^2 \right) \nonumber\\
&+ \lambda_2(t) \left( \| \theta_{c}(t) \|^2-d_c^2 \right),
\end{align}
where $ \lambda_1(t), \lambda_2(t)$ are the dual variables.

\vspace{-2ex}
Then we set the first-order partial derivative $L_j( \theta_c(t), \lambda_1\\ (t), \lambda_2 (t), \eta_c(t) ) $ with respect to $\theta_c(t)$ as zero, i.e.,
\begin{align}\label{equ:Lagrangian-first-order-derivative}
& \frac{\partial L_j(\theta_c(t), \lambda_1(t), \lambda_2(t), \eta_c(t) )}{\partial \theta_c(t)} = 2 \big( \varphi_{c,j}(t) \varphi_{c,j}^T(t)  \nonumber \\
   &  + \sigma_c^2 P_c^{-1}(\eta_c(t)) +\, \lambda_{1}(t) E_{j-1}(t+1) E_{j-1}^T(t+1)  \nonumber \\
   &  + \lambda_{2}(t) I_{n_c}  \big) \theta_c(t) - 2 \varphi_{c,j}(t) y_{c,j}(t) \nonumber \\
   &  + 2 \lambda_1(t) u_{j-1}(t) E_{j-1}(t+1)= 0,
\end{align}
which yields that
\begin{align}\label{equ:Lagrangian-first-order-derivative-closedform}
& \hat{\theta}_{c,j}( \lambda_1(t), \lambda_2(t), \eta_c(t) ) = \left( \varphi_{c,j}(t) \varphi_{c,j}^T(t)  + \sigma_c^2 P_c^{-1}(\eta_c(t)) \right. \nonumber \\
   &   \left. +\, \lambda_{1}(t) E_{j-1}(t+1) E_{j-1}^T(t+1) + \lambda_{2}(t) I_{n_c}  \right)^{-1} \nonumber \\
   &  \times \left( \varphi_{c,j}(t) y_{c,j}(t) - \lambda_{1}(t) u_{j-1}(t) E_{j-1}(t+1) \right).
\end{align}

Substituting \eqref{equ:Lagrangian-first-order-derivative-closedform} into \eqref{equ:Lagrangian-eg}, the maximization of the Lagrange dual function associated with the constrained optimization problem \eqref{equ:estimator-RLS-controller} becomes
\begin{align} \label{equ:dual-function-maximization}
& \hat{\lambda}_{1,j}(t),\hat{\lambda}_{2,j}(t)= \argmax_{\lambda_1(t),\lambda_2(t) \geq0 }  \big( y_{c,j}(t) - \varphi_{c,j}^T(t) \nonumber \\
& \times \hat{\theta}_{c,j}(\lambda_1(t), \lambda_2(t),\eta_c(t) ) \big)^2 \nonumber\\
& + \sigma_c^2 \big( \hat{\theta}_{c,j} ( \lambda_1(t),\lambda_2(t), \eta_c(t) ) \big)^T P_c^{-1}(\eta_c(t)) \nonumber \\
& \times \hat{\theta}_{c,j}( \lambda_1(t),\lambda_2(t), \eta_c(t) ) + \lambda_1(t) \big( | E_{j-1}^T(t+1) \nonumber\\
& \times \hat{\theta}_{c,j}( \lambda_1(t), \lambda_2(t), \eta_c(t) ) + u_{j-1}(t) |^2-d_u^2 \big) \nonumber\\
& + \lambda_2(t) \big( \| \hat{\theta}_{c,j}( \lambda_1(t), \lambda_2(t), \eta_c(t) ) \|^2-d_c^2 \big),
\end{align}
where $ \hat{\lambda}_{1,j}(t),\hat{\lambda}_{2,j}(t)$ are the estimates of the dual variables $ \lambda_1(t), \lambda_2(t)$ at iteration $j$ and time $t$, respectively.

Finally, with \eqref{equ:dual-function-maximization} and \eqref{equ:Lagrangian-first-order-derivative-closedform}, for each iteration $j=1,\cdots,N_e$ and time $t=1, \cdots, N_d$, we can obtain a suboptimal solution of the constrained optimization problem \eqref{equ:estimator-RLS-controller} with the closed-form expression as follows:
\begin{align}\label{equ:constrained-estimator-RLS-closedform-controller}
   & \hat{\theta}_{c,j}(\eta_c(t)) =\big( \varphi_{c,j}(t) \varphi_{c,j}^T(t)  + \sigma_c^2 P_c^{-1}(\eta_c(t)) \nonumber \\
   &  +\,\hat{ \lambda}_{1,j}(t) E_{j-1}(t+1) E_{j-1}^T(t+1) + \hat{\lambda}_{2,j}(t) I_{n_c}  \big)^{-1} \nonumber \\
   &  \times \left( \varphi_{c,j}(t) y_{c,j}(t) - \hat{\lambda}_{1,j}(t) u_{j-1}(t) E_{j-1}(t+1) \right).
\end{align}
In the second step,  for each iteration $j=1,\cdots,N_e$ and time $t=1, \cdots, N_d$,  we show that it is possible to adapt the SURE method to estimate $\eta_c(t)$ by using the closed-form expression \eqref{equ:constrained-estimator-RLS-closedform-controller}.
First, let us recall the SURE method \eqref{equ:empirical-Bayes-eg} and let
\begin{align}
H=\Phi  \left(\Phi^T \Phi+ \sigma^2  P^{-1} (\eta) \right)^{-1} \Phi^T.
\end{align}
Then it is well-known from e.g. \cite[Sections 7.4 and 7.6]{HTF2001}  that the matrix $H$ maps the measurement output $Y$ to the predicted output $\hat Y$ for the RLS estimator $\hat{\theta}^{\TR}(\eta)$, i.e., $\hat Y=H Y$, and $H$ and $\Trace (H)$ are the so-called ``hat matrix'' and ``degrees of freedom'' of the RLS estimator $\hat{\theta}^{\TR}(\eta)$, respectively, where the latter is used to describe the model complexity of the RLS estimator $\hat{\theta}^{\TR}(\eta)$ in statistics, because $0\leq \Trace (H)\leq n$. Therefore, the SURE method \eqref{equ:empirical-Bayes-eg} can be equivalently rewritten as
\begin{align}\label{equ:empirical-Bayes-eg-dof}
   &  \hat{\eta} = \argmin \limits_{\eta \in \Omega}   || Y - \Phi \hat{\theta}^{\TR}(\eta)  ||^2 + 2 \sigma^2  \Trace  (H).
\end{align}
Then, it is interesting to note that with the definition of $\varphi_{c,j}(t)$ in \eqref{equ:tracking-model-regression},
 \eqref{equ:constrained-estimator-RLS-closedform-controller} can be rewritten as
\begin{align}\label{equ:constrained-estimator-RLS-closedform-controller-hyper-partition}
   & \hat{\theta}_{c,j}(\eta_c(t)) =\big( \varphi_{c,j}(t) \varphi_{c,j}^T(t)  + \sigma_c^2 P_c^{-1}(\eta_c(t)) \nonumber \\
   &   + \hat{\lambda}_{1,j}(t) E_{j-1}(t+1) E_{j-1}^T(t+1) \!+\! \hat{\lambda}_{2,j}(t) I_{n_c}  \big)^{-1} \varphi_{c,j}(t) \nonumber \\
   &  \times \left( y_{c,j}(t) - \hat{\lambda}_{1,j}(t) (\hat{b}_{1,j}^{\TR}(t+1))^{-1} u_{j-1}(t) \right).
\end{align}
Next, we denote the prediction of $y_{c,j}(t)$ by $\hat y_{c,j}(t)$ and then we have
\begin{subequations}
\begin{align}
& \hat y_{c,j}(t) = H_{c,j}(t) \big( y_{c,j}(t) - \hat{\lambda}_{1,j}(t) (\hat{b}_{1,j}^{\TR}(t+1))^{-1} \nonumber \\
& \qquad\;\;\, \times u_{j-1}(t) \big), \\
& H_{c,j}(t) = \varphi_{c,j}^T(t) \big( \varphi_{c,j}(t) \varphi_{c,j}^T(t)  + \sigma_c^2 P_c^{-1}(\eta_c(t))  \nonumber \\
&  + \hat{\lambda}_{1,j}(t) E_{j-1}(t+1) E_{j-1}^T(t+1) + \hat{\lambda}_{2,j}(t) I_{n_c}  \big)^{-1} \nonumber  \\
& \times \varphi_{c,j}(t).
\end{align}
\end{subequations}
Now it is clear to see that $H_{c,j}(t)$ maps $y_{c,j}(t) - \hat{\lambda}_{1,j}(t) (\hat{b}_{1,j}^{\TR}(t+1))^{-1} u_{j-1}(t) $ to $\hat y_{c,j}(t)$ and moreover, $ 0 \leq \Trace (H_{c,j}(t)) = H_{c,j}(t) \leq  1 $. As a result, the hyperparameter estimate $ \hat{\eta}_{c,j}(t)$ of $\eta$ can be obtained by
\begin{align}\label{equ:hyper-parameter-estimate-controller}
   &  \hat{\eta}_{c,j}(t) = \argmin \limits_{\eta_c(t) \in \Omega} \left( y_{c,j}(t) - \varphi_{c,j}^T(t) \hat{\theta}_{c,j}(\eta_c(t))  \right)^2 \nonumber\\
   &\qquad\;\;\, + 2 \sigma_c^2  H_{c,j}(t) .
\end{align}

\subsection{Summary}\label{subsec3-3}
The proposed kernel-based regularized ILC (KRILC) method can be summarized in Algorithm 1.

\begin{tabular}{p{8cm}}
  \hline
  \multicolumn{1}{l}{\textbf{Algorithm 1:} KRILC}\\
  \hline
\textbf{Require: }Initialize the input-output data with zero values. \\
\textbf{For} $j=1, \cdots, N_e$ \\
 \quad \textbf{For} $t=1, \cdots, N_d$ \\
 \quad \quad Compute $\hat{\theta}_{m,j}^{\TR}(t+1)$ by (\ref{equ:estimator-RLS-model}) and (\ref{equ:hyper-parameter-estimate-model}). \\
 \quad \quad Compute $ \hat{\eta}_{c,j}(t)$ by (\ref{equ:hyper-parameter-estimate-controller}). \\
 \quad \quad Compute $\hat{\theta}_{c,j}^{\TR}(t)$ by (\ref{equ:estimator-RLS-controller}). \\
 \quad \quad Compute $u_j(t)$ by (\ref{equ:ILC-law-applied}) and apply it to the time-\\ \quad \quad varying system (\ref{equ:system}). \\
\quad \textbf{EndFor} \\
\textbf{EndFor} \\
 \hline
\end{tabular}


\begin{remark}\label{rmk: order-hyperparameters}
It is worth to stress that the model orders $n_a$, $n_b$ and $n_c$ can be chosen to be sufficiently large. The reason is that for KRM, the model complexity is not governed by them but by the corresponding hyper-parameters, which can be tuned in a continuous and automatic way, e.g., \cite{COL2012}. In addition, the unknown variances $\sigma^2$ and $\sigma_c^2$ are estimated by treating them as additional hyper-parameters, e.g., \cite{MacKayDJ}. 
\end{remark}

\begin{remark}
There are different kinds of models to describe linear time-varying systems, such as the state-space model, e.g., \cite{LLK2000, GN2006, NB2011}, and the ARX model, e.g., \cite{J2018, JX2013,HLS2010}, which is considered in this paper. Accordingly, there are two related classes of methods to handle the problems of model estimation and controller design in ILC. The first class of methods, e.g., \cite{LLK2000, GN2006, NB2011}, first convert the state-space model estimation problem into the estimation problem of the zero-state impulse response matrix of the state-space model, and then update the zero-state impulse response matrix and the parameters of learning controller once per iteration. Different from the first class of methods, the second class of methods, e.g., \cite{J2018, JX2013,HLS2010}, update parameters of the ARX model and the learning controller at each time per iteration based on the data up to current time and iteration, as done in this paper. Moreover, there are many successful applications for the two classes of methods. For example, the first class of methods have been applied in chemical batch reactors, e.g., \cite{LLK2000}, and the second class of methods have been applied in robots, e.g. \cite{ZLXF2014, J2018}, and linear motor systems, e.g., \cite{YHX2018}.
\end{remark}



\section{Ultimate Boundedness of the Tracking Error}\label{sec4}
In this section, we show that  the tracking error \eqref{equ:error} for the proposed KRILC method is ultimately bounded in the iteration domain. To show this result, we need the following assumptions.

\begin{assumption}\label{asmp:bounded-tracking}
Assume that for each time $t=1, \cdots, N_d$, the tracking error $e_1(t)$ is bounded.
\end{assumption}

\begin{assumption}\label{asmp:bounded-reference}
Assume that for each time $t=1, \cdots, N_d$, the reference output $y_d(t)$ is bounded above by a constant $d_r>0$, i.e. $| y_d(t) | \leq d_r$.
\end{assumption}

\begin{assumption}\label{asmp:bounded-controller}
Assume that for each iteration $j=1,\cdots,N_e$ and time $t=1,\cdots,N_d$,  $u_j(t)$, $v_j(t),\theta_c(t)$ are bounded above by the constants $d_u,d_v,d_c>0$, respectively, i.e., $| u_j(t) | \leq d_u$, $| v_j(t) | \leq d_v$, and $\| \theta_c(t) \| \leq d_{c}$.
 \end{assumption}

\begin{assumption}\label{asmp: BIBO-stable-system}
Assume that the linear time-varying system (\ref{equ:system}) is uniformly BIBO stable.
\end{assumption}

\begin{remark} \label{rmk: BIBO-stable-system}

Recall that the uniform BIBO stability in Definition \ref{def:BIBO-stability-eg} or \cite[Definition 27.1]{R1996} is defined for linear time-varying systems that can be described by linear state-space models in the form of \eqref{equ:state-model-eg}. Therefore, in order to justify that Assumption \ref{asmp: BIBO-stable-system} makes sense, we need to show that the time-varying system (\ref{equ:system}) indeed obtains a linear state-space model in the form of \eqref{equ:state-model-eg}. To this goal, for each iteration $j=1, \cdots, N_e$ and time $t=1, \cdots, N_d$, we let $x_{j,1}(t-1) = y_j(t-1)$, $\cdots$, $x_{j,n_a}(t-1) = y_j(t-n_a)$, which gives that
\begin{align}
   & x_{j,1}(t) = y_j(t), \nonumber \\
   & x_{j,2}(t) = y_j(t-1) = x_{j,1}(t-1), \nonumber \\
   & \vdots, \nonumber \\
   & x_{j,n_a}(t) = y_j(t-n_a+1) = x_{j,n_a-1}(t-1). \nonumber
\end{align}
Then the time-varying system \eqref{equ:system} has the following linear state-space model:
\begin{subequations}\label{equ:state-output-model}
\begin{align}
   & x_j (t+1) =\mathcal{ A}(t) x_j(t) + \mathcal{B}(t) \mathfrak{u}_j (t), \\
   & y_j (t) = \mathcal{C}(t) x_j (t),
\end{align}
\end{subequations}
where $ \mathcal{A}(t) \in \mathbb {R}^{n_a \times n_a} $, $\mathcal{B}(t) \in \mathbb{R}^{n_a \times (n_b+1)}$,  $\mathcal{C}(t) \in \mathbb{R}^{n_a}$,
\begin{align}
   & x_j (t) = [ x_{j,1}(t), \cdots, x_{j,n_a}(t) ]^T\in\mathbb {R}^{n_a}, \nonumber \\
   &\mathfrak{u}_j (t) = [ u_j (t), \cdots, u_j (t-n_b+1), v_j(t+1) ]^T \in \mathbb{R}^{n_b+1}, \nonumber
   \end{align}
\begin{align}
   & \mathcal{A}(t) \!=\!\! \left(\!
              \begin{array}{ccccc}
                -a_1(t\!+\!1) & \cdots  & -a_{n_a-1}(t\!+\!1) & -a_{n_a}(t\!+\!1)\\
                        1           & \cdots              & 0   & 0 \\
                        0           & \cdots              & 0   & 0 \\
                \vdots    & \ddots             & \vdots  & \vdots \\
                        0          & \cdots              & 1 & 0 \\
              \end{array}
           \! \right)\!, \nonumber \\
   & \mathcal{B}(t) = \left(
              \begin{array}{cccc}
                b_1(t+1)  & \cdots & b_{n_b}(t+1) & 1 \\
                        0          & \cdots & 0               & 0   \\
                \vdots   & \ddots & \vdots       & \vdots  \\
                        0           & \cdots  & 0              & 0  \\
              \end{array}
            \right), \nonumber \\
   & \mathcal{C} (t) = [ 1, 0, \cdots, 0 ]. \nonumber
\end{align}

 \end{remark}

\begin{theorem} \label{thrm:convergence}
Consider the time-varying system (\ref{equ:system}), and let it be controlled by the KRILC as summarized in Algorithm 1. Suppose that Assumptions \ref{asmp:bounded-tracking}-\ref{asmp: BIBO-stable-system} are satisfied. Then for each time $t =1, \cdots, N_d $, there exist two finite constants $d_{g,u}$ and $d_{g,v}$ with $d_{g,u}^2 +d_{g,v}^2 = d_g^2$ such that the tracking error \eqref{equ:error} is ultimately bounded, i.e.,
\begin{align}\label{equ:tracking-inequality-final}
   & \lim \limits_{j \rightarrow \infty} | e_j(t+1) | \leq  \Bigg( \left(\sqrt{n_b} d_{g,u} d_u + d_{g,v} d_v \right) \nonumber \\
   &   \times \left(2 d_{g,u} d_c  \sqrt{\frac{n_b n_c (n_c^2-1)}{2}}  +1 \right)  +d_r \Bigg) \nonumber \\
   & \; \Bigg/ \left(1- d_{g,u} d_c \sqrt{\frac{n_b n_c (n_c+1)}{2}}\right),
\end{align}
if it holds that
  \begin{equation}\label{equ:condition-theorem1}
  \begin{aligned}
     d_{g,u} d_c \sqrt{\frac{n_b n_c (n_c+1)}{2}} < 1.
  \end{aligned}
  \end{equation}
\end{theorem}

\begin{remark}\label{rmk: conservative-conditions}
The condition \eqref{equ:condition-theorem1} is sufficient to guarantee that the tracking error \eqref{equ:error} is ultimately bounded. In our numerical simulations, we observe that the tracking error \eqref{equ:error} can still be ultimately bounded even if the condition \eqref{equ:condition-theorem1} is violated, indicating that the condition \eqref{equ:condition-theorem1} is kind of conservative. It can be seen from \eqref{equ:tracking-inequality-final} that as $d_{g,u}$ decreases, the ultimate bound of the tracking error \eqref{equ:error} will decrease. Moreover, the constants $d_{g,v}$, $d_v$, $d_u$ and $d_r$ can be understood in the similar way with respect to $d_{g,u}$.
\end{remark}

For all parameters in this paper, we give a summary for some guidelines of their tuning and determination as shown in Tables \ref{tbl:guideline-tuned} and \ref{tbl:guideline-determined}. Note that these parameters can be divided into two classes:
\begin{itemize}
  \item The first class of parameters include $\eta_m(t)$, $\eta_c(t)$, $\sigma^2$ and $\sigma_c^2$, and they are tuned automatically based on the data.
  \item The second class of parameters include $d_u$, $d_v$, $d_r$, $d_{g,u}$, $n_a$, $n_b$, $n_c$ and $d_c$, and they are determined by the underlying system and controller, but not tuned by the users.
\end{itemize}

\begin{table}[tbhp]
\centering
\caption{The parameters that are tuned automatically based on the data.}\label{tbl:guideline-tuned}
\begin{tabular}{p{0.9cm}p{6.5cm}}
   \hline
   Notation & Guideline \\
   \hline
   $\eta_m(t)$  & tuned according to \eqref{equ:hyper-parameter-estimate-model}.  \\
   $\eta_c(t)$   & tuned according to \eqref{equ:hyper-parameter-estimate-controller}. \\
   $\sigma^2, \sigma_c^2$  & treated as additional hyper-parameters of $\eta_m(t)$ and $\eta_c(t)$, respectively. \\
   \hline
\end{tabular}
\end{table}

\begin{table}[tbhp]
\centering
\caption{The parameters that are determined by the underlying system and controller (they are not tuned by users).} \label{tbl:guideline-determined}
\begin{tabular}{p{0.9cm}p{6.5cm}}
   \hline
   Notation & Guideline \\
   \hline
   $d_u$                          & determined by the upper bound of input $u_j(t)$ of system \eqref{equ:system}. \\
   $d_v$                          & determined by the upper bound of measurement noise $v_j(t)$. \\
   $d_r$                          & determined by the upper bound of reference output $y_d(t)$. \\
   $d_{g,u}$                   & determined by the upper bound of zero-state response of system \eqref{equ:system} in the case where it is assumed that $v_j(t)=0$.  \\
   $n_a$, $n_b$, $n_c$  & determined by system \eqref{equ:system} and learning controller \eqref{equ:ILC-law} and no bias-variance tradeoff is included in their selection, and just need to be selected sufficiently large to capture their dynamics. \\
   $d_c$                          & determined by the condition \eqref{equ:condition-theorem1} as long as $d_{g,u}$, $n_b$, $n_c$ are fixed. \\
   \hline
\end{tabular}
\end{table}

\section{Numerical Simulation}\label{sec5}
In this section, we run numerical simulations to show the efficacy of the proposed KRILC method. Note that in the simulations, most of the quantities have no units, such as the fit and iteration number, except the time, and its unit is second.

\subsection{KRILC: a Numerical Example from \cite{ShenZWC16}}\label{subsec5-example1}

We first consider the numerical example studied in \cite[p. 363--364]{ShenZWC16}.

\emph{1) Test System and Reference Output: }
The linear time-varying system of this numerical example can be written in the form of (\ref{equ:system}):
\begin{align}\label{equ:system1-simulation}
  & \left(1 + a_1(t) q^{-1} + a_2(t) q^{-2} \right) y_j(t) = \left(  b_1(t) q^{-1} \right. \nonumber \\
  & \left.+ b_2(t) q^{-2} \right) u_j(t) +v_j(t),
\end{align}
where $a_1(t) = 1.2$, $a_2(t) = -0.35$, $b_1(t) = 1 +0.1 \cos (t) + 0.03 \sin (t)$ and $b_2(t) = 0.1 \sin(t) -0.05 \cos (t) -0.521$, and $v_j(t)$ is the measurement noise we added artificially.

The reference output, as suggested in  \cite{ShenZWC16}, is
\begin{equation}\label{equ:output-reference-system1}
\begin{aligned}
    y_d(t)=\sin \left(\frac{2 \pi t}{50} \right) + \sin \left(\frac{2 \pi t}{5} \right).
\end{aligned}
\end{equation}

 \begin{figure}[h]
 \centering
 \includegraphics [width=7cm]{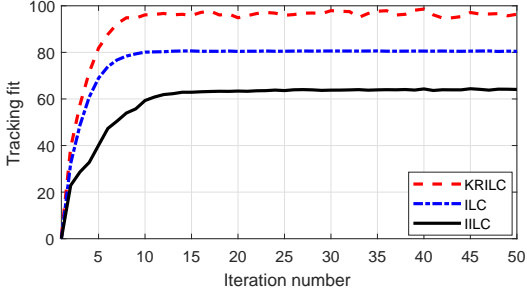}
 \vspace{-1.5ex}
 \caption{The tracking fits of system \eqref{equ:system1-simulation} along the iteration axis for the KRILC, data-driven ILC in \cite{YuX22TC} and inversion-based ILC in \cite{dRO2019}, which are labeled by KRILC, ILC and IILC, respectively.}\label{fig:control-fit-system1}
\end{figure}

\begin{figure}[h]
 \centering
 \includegraphics [width=7cm]{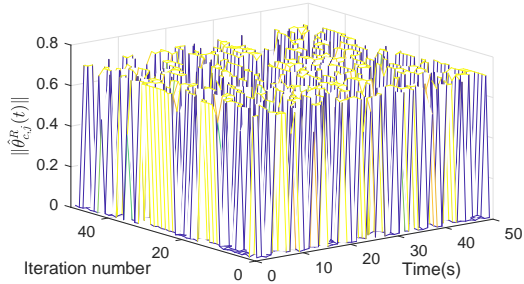}
 \vspace{-1.5ex}
 \caption{The magnitudes of $ \| \hat{\theta}_{c,j}^{\TR}(t) \| $ along the iteration and time axes for the KRILC, which shows that the constraint $ \| \hat{\theta}_{c,j}^{\TR}(t) \|  \leq 0.7$ is satisfied.}
 \label{fig:norm-theta-system1}
\end{figure}

\emph{2) Simulation Setup: }
We set $N_e=50$, $N_d=50$. For each time $t=1, \cdots, 50$, we first use the method in \cite{YuX22TC} to design a stable controller and apply it to system (\ref{equ:system1-simulation}), and collect the initial experiment of data. Then, for each iteration $j=1, \cdots, 50$ and time $t=1, \cdots, 50$, we apply the proposed KRILC and in particular, we choose $n_a=n_b=n_c=10$, $d_u=2$, $d_c=0.7$, the RLS model and controller estimators with the DI kernel; we take the noise variances $\sigma^2$ and $\sigma_c^2$ as additional hyper-parameters of $\eta_m(t)$ and $\eta_c(t)$, respectively; we compute the input $u_j(t)$ by (\ref{equ:ILC-law-applied}) and apply it to system (\ref{equ:system1-simulation}), and collect the data $u_j(t)$ and $y_j(t)$ where $y_j(t)$ is perturbed by the zero mean noise $v_j(t)$ with $\sigma^2 = 0.01$ and $d_v=0.05$. For comparisons, we also consider the data-driven ILC in \cite{YuX22TC} and the inversion-based ILC in \cite{dRO2019}. Moreover, to show the estimate of $\theta_m(t)$, we set $N_e=500$, and for comparison, we also consider the LS model estimator, i.e.,  (\ref{equ:estimator-RLS-model}) with $P_m(t)^{-1} = 0$.

\begin{figure}[h]
 \centering
 \includegraphics [width=7cm]{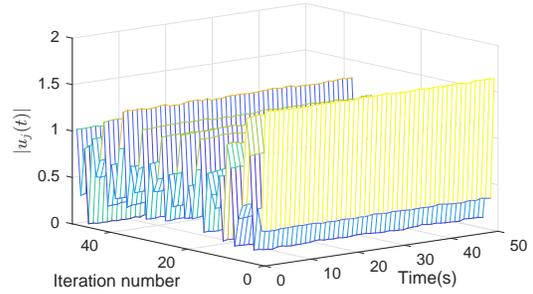}
 \vspace{-1.5ex}
 \caption{The magnitudes of $ | u_j(t) | $ along the iteration and time axes for the KRILC, which shows that the constraint $ | u_j(t) |  \leq 2$ is satisfied.}
 \label{fig:absolute-input-system1}
\end{figure}

\begin{figure}[h]
 \centering
 \includegraphics [width=7cm]{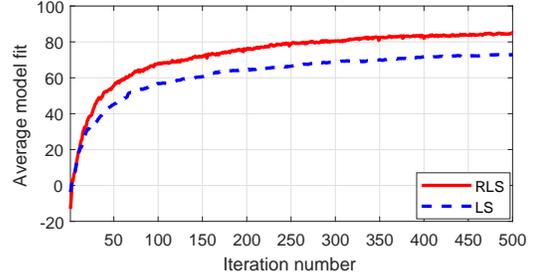}
 \vspace{-1.5ex}
 \caption{The average model fits of system \eqref{equ:system1-simulation} along the iteration axis for the the RLS model estimator (\ref{equ:estimator-RLS-model}) and the LS model estimator, which are labeled by RLS and LS, respectively.}\label{fig:model-fit-system1}
\end{figure}

The learning controller for the method in \cite{YuX22TC} is given by
\begin{subequations}\label{equ:ILC}
\begin{align}
    & u_{j}(t) = u_{j-1}(t) + \xi_{j-1}^T(t+1) \theta_{j}(t), \\
    & \xi_{j-1}(t+1) = \left[ - e_{j-1}(t+1), \Delta e_{j-1}(t+1), \cdots, \right. \nonumber \\
    & \qquad\qquad\qquad \left. \Delta e_{j-l_{\theta}+1} \right]^T, \\
    & \theta_{j+1}(t) = \theta_{j}(t) + \eta_{\theta}  \Big( \hat{\psi}_j(t) e_{j-1}(t+1) - \mu_{\theta} \xi_{j-1}^T(t+1) \nonumber \\
    & \quad \times \theta_{j}(t) \Big) \xi_{j-1}(t+1) \Big/ \left( \big( \mu_{\theta} +  \hat{\psi}_j^2(t) \big) \| \xi_{j-1}(t+1) \|^2 \right)\!, \\
    & \hat{\psi}_{j}(t) = \hat{\psi}_{j-1}(t) + \eta_{\psi} \left( \Delta y_j(t) - \hat{\psi}_{j-1}(t) \Delta u_{j}(t-1)  \right) \nonumber \\
    & \qquad\; \times \Delta u_{j}(t-1) \Big/ \left( \mu_{\psi} + \Delta u_{j}^2(t-1)  \right),
\end{align}
\end{subequations}
 where $\Delta$ denotes the backward difference operator of a variable in the iteration domain, e.g., $\Delta e_{j}(t+1) = e_{j}(t+1)- e_{j-1}(t+1)$, $l_{\theta} \in \mathbb{N}$ denotes the order of the learning controller \eqref{equ:ILC}, $\eta_{\theta} \in (0,1]$, $\mu_{\theta} >0$, $\eta_{\psi} \in (0,1]$, and $\mu_{\psi} >0$ are the parameters that are needed to be tuned in an ad-hoc way. Here we choose $l_{\theta}=3$, $\eta_{\theta}=0.1$, $\mu_{\theta} = 0.5$, $\eta_{\psi}=1$, and $\mu_{\psi}=1$. \\
 The learning controller for the method in \cite{dRO2019} is given by
\begin{subequations}\label{equ:IILC}
\begin{align}
    & \mathcal{U}_{j} \!=\!\left\{ \!
            \begin{aligned}
                &  \mathcal{U}_{j-1}  + \rho(| \mathcal{Y}_{j-1} |) \frac{\mathcal{U}_{j-1} }{\mathcal{Y}_{j-1}} \mathcal{E}_{j-1},\!  &  \mathcal{Y}_{j-1} \neq 0, \\
                &  \mathcal{U}_{j-1},
                &  \mathcal{Y}_{j-1} = 0,
             \end{aligned} \right. \\
     &  \rho(| \mathcal{Y}_{j-1} |) \!=\!   \left\{\!
              \begin{aligned}
                  &  1,  & | \mathcal{Y}_{j-1} | > \gamma, \\
                  & \frac{1}{2} \left( 1 - \cos \left( \frac{\pi}{\gamma}  | \mathcal{Y}_{j-1} | \right) \right), &  | \mathcal{Y}_{j-1} | \leq \gamma,
               \end{aligned} \right.
\end{align}
\end{subequations}
where $\mathcal{U}_{j}=\mathcal{U}_{j}(k) = \mathcal{F}\{ u_j(t) \}$, $\mathcal{Y}_{j}=\mathcal{Y}_{j}(k) = \mathcal{F}\{ y_j(t) \}$, $\mathcal{E}_{j}=\mathcal{E}_{j}(k) = \mathcal{F}\{ e_j(t) \}$, $k=0, 1, \cdots, N-1$, $N \in \mathbb{N}$ is a constant, $\mathcal{F}\{\cdot \}$ denotes the $N$-point Discrete Fourier Transform, $\gamma >0$ is the parameter that is needed to be tuned in an ad-hoc way. Here we choose $\gamma=0.9$.

\emph{3) Simulation Results and Discussions: }
To assess the aforementioned three ILC methods, for each iteration $j\in\mathbb N$, we calculate the tracking fit by
\begin{subequations}\label{equ:fit-control}
\begin{align}
 & \text{fit}_j^c=100 \left( 1 - \sqrt{ \frac{ \sum \limits_{t=1}^{N_d} \left( y_d(t) - y_j(t) \right)^2 }{ \sum \limits_{t=1}^{N_d} \left( y_d(t) - \bar{y}_d \right)^2 } }  \right)\!, \\
 & \bar{y}_d =\frac{1}{N_d} \sum \limits_{t=1}^{N_d} y_d(t).
\end{align}
\end{subequations}

The results are depicted in Fig. \ref{fig:control-fit-system1}, which shows that in contrast with the two ILC methods in \cite{YuX22TC} and \cite{dRO2019}, KRILC has faster convergence speed and better accuracy in terms of the tracking fit. Moreover, for KRILC, and for each iteration $j=1, \cdots, 50$ and time $t=1, \cdots, 50$, the magnitudes of $ \| \hat{\theta}_{c,j}^{\TR}(t) \| $ and $ | u_j(t) | $ are depicted in Figs. \ref{fig:norm-theta-system1} and \ref{fig:absolute-input-system1}, respectively, which show that the two constraints $ \| \hat{\theta}_{c,j}^{\TR}(t) \|  \leq 0.7$ and $ | u_j(t) |  \leq 2$ are both satisfied.

Moreover, to assess the two model estimators, for each iteration $j=1, \cdots, 500$ and time $t=1, \cdots, 50$, we calculate the model fit \cite{Ljung1995, COL2012} by
\begin{subequations}\label{equ:fit-model}
\begin{align}
 & \text{fit}_j(t) = 100 \left( 1 - \sqrt{ \frac{ \sum \limits_{l=1}^{n_a+n_b} \left( \theta_m^l(t) - \hat\theta_{m,j}^{\TR,l}(t) \right)^2 }{\sum \limits_{l=1}^{n_a+n_b} \left( \theta_m^l(t) - \bar{\theta}_m^l(t) \right)^2 } }  \right), \\
 & \bar{\theta}_m(t) =\frac{1}{n_a+n_b}\sum \limits_{l=1}^{n_a+n_b}\theta_m^l(t),
\end{align}
\end{subequations}
where $\theta_m^l(t)$ and $\hat\theta_{m,j}^{\TR,l}(t)$ denote the $l$th element of $\theta_m(t)$ and $\hat\theta_{m,j}^{\TR}(t)$, respectively. Then for each iteration $j=1, \cdots, 500$, we have 50 model fits corresponding to the 50 time instants of model parameters and finally, we calculate the average model fit. In this way, for the 500 iterations, we have 500 average model fits, as shown in Fig. \ref{fig:model-fit-system1}, which shows that the average model fit increases as the number of iterations increases and in contrast with the LS model estimator, the RLS model estimator \eqref{equ:estimator-RLS-model} has better average model fits.

\begin{figure}[h]
 \centering
 \includegraphics[width=5cm]{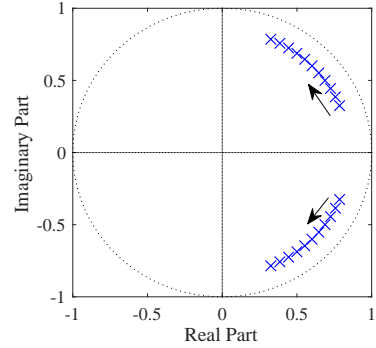}
 \caption{Time-varying trajectories for one pair of complex-conjugate poles: $\times$ denotes the location of the pole whose moving direction is denoted by $\rightarrow$.}\label{fig:pole-zero}
\end{figure}

\begin{figure}[h]
 \centering
 \includegraphics [width=6.5cm]{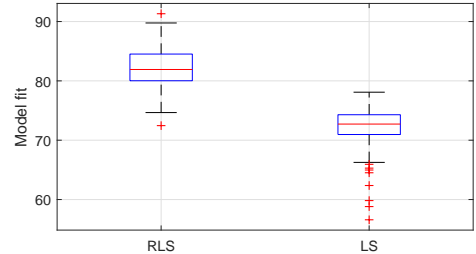}
 \vspace{-1.5ex}
 \caption{Box plot of the 500 average model fits for the RLS model estimator (\ref{equ:estimator-RLS-model}) and the LS model estimator, which are labeled by RLS and LS, respectively.}\label{fig:model-fit}
\end{figure}

\begin{remark}
It is worth to note that the definitions of tracking fit \eqref{equ:fit-control} and model fit \eqref{equ:fit-model} are normalized measure of the tracking and model estimation performances, and widely used in the literature, e.g., \cite{Ljung1995, COL2012,PDCDL2014, PCCDL2022}, and also in the Matlab system identification toolbox (the function \textit{compare}). The maximum value of the two fits is 100, which means a perfect match between the true one and the estimated one. As the fits become smaller, then the tracking and model estimation performances would become worse. Once the fits become negative, it can be regarded that the tracking and model estimation performances are very bad.
\end{remark}

\subsection{KRILC: Monte Carlo Simulations}\label{subsec5-KRILC}

Now we run Monte Carlo simulations to test the proposed KRILC in terms of both the model estimation and controller design. Here we run Monte Carlo simulations due to the reason as follows. It is known that there are many data-driven ILC methods, e.g., \cite{YuX22TC, HCG2016, dRO2019}, however, all of these methods have some parameters that have to be tuned manually, which makes Monte Carlo simulations impossible (Monte Carlo simulations are the common way to assess and to compare the performances of related methods). Different from these existing data-driven ILC methods, the proposed one has all of its parameters tuned in an automatic way, which gives the users great convenience for Monte Carlo simulations.

\subsubsection{Model Estimation}\label{subsec5-ModelEstimation}

\emph{1) Test Systems: }
We generate 500 test systems in the form of \eqref{equ:system} in the following way:
\begin{itemize}
  \item Generate a 10th order discrete-time linear time invariant (LTI) system with its all poles and zeros inside a circle with radius 0.95 and center at the origin using the method in \cite{COL2012}.

\item For each time $t=1, \cdots, N_d$, we construct a linear time-varying system based on the LTI system generated above as follows: for each complex pole or zero (denoted by $s_0$) of the LTI system, we change it to be a time-varying one (denoted by $s(t)$) according to
     \begin{subequations}\label{equ:poles-zeros-change}
      \begin{align}
         & s(t) \!=\! | s_0 | \left( \cos \left(\text{sgn}(a_0) \frac{\pi t}{4 N_d} \!+\! a_0 \right) \!\right.\nonumber \\
         & \quad\;\;+\left.  \!i \sin \left(\text{sgn}(a_0) \frac{\pi t}{4 N_d} \!+\! a_0 \right) \right), \\
         & a_0 = \arctan \left(\frac{ \text{Im}(s_0) }{ \text{Re}(s_0) } \right), \quad t=1, \cdots, N_d,
      \end{align}
      \end{subequations}
where $ \text{Im}(s_0)$ and $ \text{Re}(s_0) $ are the imaginary and real parts of $s_0$, and $\text{sgn}(a_0)$ denotes the sign of $a_0$. One example for a pair of generated time-varying poles is shown in Fig. \ref{fig:pole-zero}. For each time $t=1, \cdots, N_d$, we save the generated system as \texttt{dst}.

  \item For each time $t=1, \cdots, N_d$, generate an output error model \texttt{ds} based on the command \texttt{ds=idpoly(dst) } in MATLAB, and then change \texttt{ds} to an ARX model using the command \texttt{ds.d=ds.f}.
\end{itemize}

\begin{figure*}[tbhp]
 \centering
  \includegraphics [width=14.5cm]{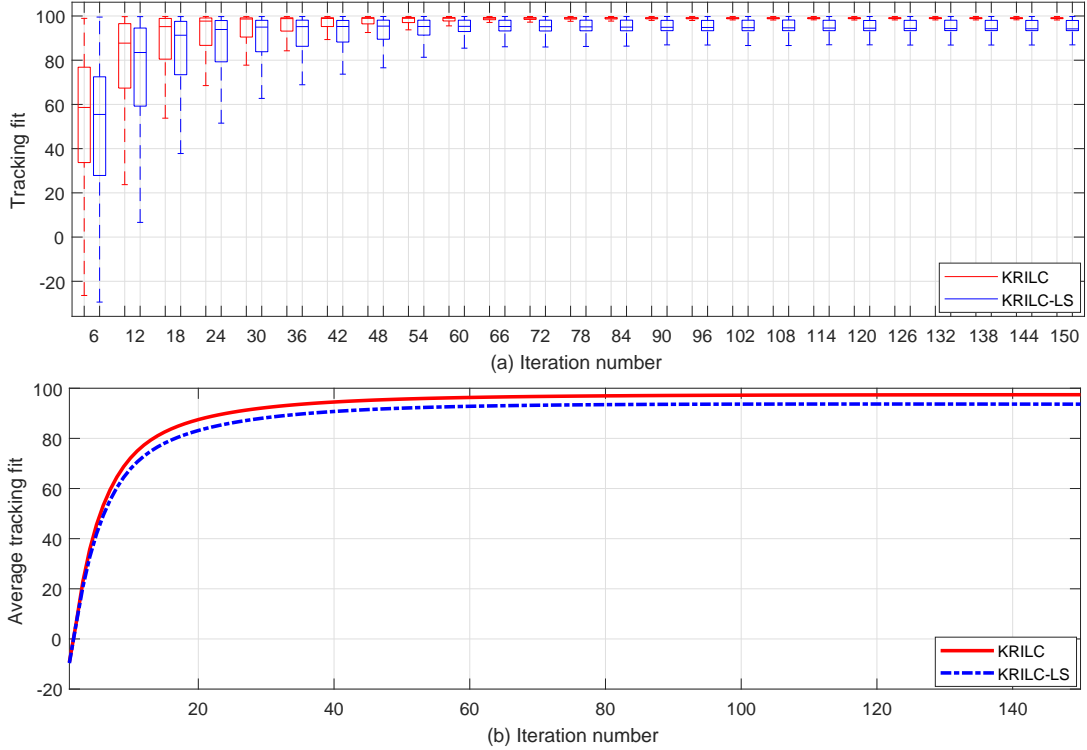}
 \vspace{-1.5ex}
 \caption{Profile of the tracking fits and average tracking fits for the 500 test systems along the iteration axis for the KRILC and its special case where where the LS controller estimator (i.e., (\ref{equ:estimator-RLS-controller}) with $P_c^{-1}(t) = 0$) is used. The two ILC methods are labeled by KRILC and KRILC-LS, respectively. Panel (a): Box plot of the 500 tracking fits, where the outliers are hidden for a clear presentation. Panel (b): Average tracking fits.}\label{fig:control-fit}
\end{figure*}

\emph{2) Test Data-bank: }
For each test system, we generate the test input as a white noise with zero mean and unit variance to simulate each test system, and the obtained output is perturbed by a noise with zero mean and unit variance. 
Following the above procedure, for each test system, we generate $200$ iterations of input-output data ($N_e=200$) with $N_d=200$. 

\emph{3) Simulation Setup: }
We choose $n_a=n_b=20$, the RLS model estimator with the DI kernel. We take the noise variance $\sigma^2$ as an additional hyper-parameter of $\eta_m(t)$. For comparison, we also consider the LS model estimator, i.e.,  (\ref{equ:estimator-RLS-model}) with $P_m(t)^{-1} = 0$. 

\emph{4) Simulation Results and Discussions: }
To assess the aforementioned two model estimators, we calculate the model fit \eqref{equ:fit-model}. Then for each test system, we have 200 model fits corresponding to the 200 time instants of model parameters and finally, we calculate the average model fit. In this way, for the 500 test systems, we have 500 average model fits, as shown in Fig. \ref{fig:model-fit}, which shows that in contrast with the LS model estimator, the RLS estimator (\ref{equ:estimator-RLS-model}) has better average model fits and also better robustness (the distribution of the average model fits is more compact).

\subsubsection{Controller Design}\label{subsec5-ControllerDesign}
%

\emph{1) Test Systems and Reference Output: }
We generate 500 test systems in the form of (\ref{equ:system}) in the following way:
\begin{itemize}
  \item Generate a system by following the procedure in 1) of Section \ref{subsec5-ModelEstimation}, and for each time $t=1, \cdots, N_d$, save the generated system as \texttt{ds} if the two conditions $\sum\limits_{i=0}^{t-1} \| G_u (t,i) \| \leq d_{g,u}$ with $d_{g,u}=5$ (c.f. \eqref{equ:A-output-model-response-ahead-partition} and \eqref{equ:A-bounded-unit-response-partioned}) and $\max\{| p(t) | \}_{t=1}^{N_d} - \min \{ | p(t) | \}_{t=1}^{N_d}  <=0.001$ with $ \max\{ | p(t)| \}_{t=1}^{N_d} >0.7 $ are satisfied, if otherwise, repeat the procedure. Here $p(t)$ is the pole with the maximum absolute magnitude of residues (see e.g. \cite[p. 180-181]{O2010}) evaluated at the poles of the generated system \texttt{ds} at time $t$, and $\max \{  \cdot\}, \min \{  \cdot\}$ denote the maximum and minimum elements of a set, respectively.
\end{itemize}

The reference output is given by
\begin{equation}\label{equ:output-reference}
\begin{aligned}
   y_d(t)=\frac{1}{8} 10^{-6}  t^3 \left(7-0.03 t \right).
\end{aligned}
\end{equation}

\emph{2) Simulation Setup: }
For each test system, we set $N_e =150$ and $N_d=200$, and for each time $t=1, \cdots, 200$, we first use the method in \cite{YuX22TC} to design a stable controller and apply it to each test system, and collect the initial experiment of data. Then for each test system, and for each iteration $j=1, \cdots, 150$ and time $t=1, \cdots, 200$, we apply the proposed KRILC and in particular, we choose $n_a=n_b=20$, $n_c=10$, $d_u=15$, $d_c=0.3$, and the RLS model and controller estimators with the DI kernel; we take the noise variances $\sigma^2$ and $\sigma_c^2$ as additional hyper-parameters of $\eta_m(t)$ and $\eta_c(t)$, respectively; we compute the input $u_j(t)$ by (\ref{equ:ILC-law-applied}) and apply it to each test system, and collect the data $u_j(t)$ and $y_j(t)$ where $y_j(t)$ is perturbed by the zero mean noise $v_j(t)$ with $\sigma^2 = 0.01$ and $d_v=0.05$. Finally, for comparison, we also consider a special case of KRILC, where the LS controller estimator (i.e., (\ref{equ:estimator-RLS-controller}) with $P_c^{-1}(t) = 0$) is used and this case is denoted by KRILC-LS.




\emph{3) Simulation Results and Discussions: }
For each test system and iteration $j=1, \cdots, 150$, we use the tracking fit (\ref{equ:fit-control}) to assess the aforementioned two ILC methods. For each iteration $j=1, \cdots, 150$, the boxplot of the 500 tracking fits are shown in panel (a) of Fig. \ref{fig:control-fit}, and the average tracking fits are shown in panel (b) of Fig. \ref{fig:control-fit}, which show that in contrast with a special case of KRILC, where the LS controller estimator (i.e., (\ref{equ:estimator-RLS-controller}) with $P_c^{-1}(t) = 0$) is used, KRILC has faster convergence speed and better accuracy in terms of average tracking fits, and also better robustness (the distribution of the tracking fits is more compact). Moreover, at the 150th iteration, the average tracking fits of KRILC and KRILC-LS are 97.42 and 93.65, respectively.

\section{Conclusion}\label{sec6}
In this paper, we integrated the kernel-based regularization method into the data-driven ILC for a class of repetitive linear time-varying systems. This integration contributes to the proposed method with two characteristics that are different from the existing data-driven ILC methods, e.g., \cite{dRO2019, YuX22TC}. One is that the prior knowledge of underling system and controller can be embedded into the model estimation and controller design problems. Another one is that all of the parameters are tuned in an automatic way for the proposed method, while the existing ones have some parameters needed to be tuned in an ad-hoc way. The numerical simulation results showed that the proposed method could give faster convergence speed, better accuracy and robustness in terms of the tracking performance in comparison with the least squares (LS) method, and the data-driven ILC methods in \cite{dRO2019, YuX22TC}, illustrating the advantages of the kernel-based regularization method in handling both the model estimation and controller design problems.


\renewcommand{\thesection}{A}
\setcounter{thm}{0}
\section{Appendix}\label{secA}
This appendix contains the detailed derivations of Theorem \ref{thrm:convergence}.

First of all, according to Remark \ref{rmk: BIBO-stable-system}, Assumption \ref{asmp: BIBO-stable-system} and Lemma \ref{lmm:BIBO-response-eg}, for each iteration $j=1, \cdots, N_e$ and time $t=1, \cdots, N_d$, we have that the zero-state response of the time-varying system (\ref{equ:system}) or equivalent \eqref{equ:state-output-model} can be represented as
\begin{align} \label{equ:output-model-response}
     y_j (t) = \sum\limits_{i=0}^{t-1} G (t,i) \mathfrak{u}_j (i),
\end{align}
where $G (t,i)$ is defined in \eqref{eq:impulse response of linear time varying system} with
$I_{n_x}$ replaced by $I_{n_a}$.
Moreover, there exists a finite constant $d_g$ such that for each time $t=1,\cdots,N_d$,
\begin{align}\label{equ:A-bounded-unit-response}
   \sum\limits_{i=0}^{t-1} \| G (t,i) \| \leq d_g,
\end{align}
 and for each iteration $j=1, \cdots, N_e$ and time $t=1, \cdots, N_d$,
 \begin{align}\label{equ:BIBO-stable}
   \sup\limits_{ t \geq 1} | y_j(t) | \leq d_g \sup\limits_{ t \geq 1} \| \mathfrak{u}_j(t) \|.
\end{align}

Now we let $G(t,i)= [G_u(t,i), G_v(t,i)]$ with
\begin{align}
   & G_u (t,i) = \mathcal{C} (t) \Psi(t,i+1) \bar{\mathcal{B}}(i),   \nonumber \\
   & G_v (t,i) = \mathcal{C} (t) \Psi(t,i+1) \mathcal{C}^T (i),   \nonumber \\
   & \bar{\mathcal{B}}(t) = \left(
              \begin{array}{ccc}
                b_1(t+1)  & \cdots & b_{n_b}(t+1)  \\
                        0          & \cdots & 0                  \\
                \vdots   & \ddots & \vdots         \\
                        0           & \cdots  & 0              \\
              \end{array}
            \right), \nonumber
\end{align}
and then for each iteration $j=1, \cdots, N_e$ and time $t=1, \cdots, N_d$, according to \eqref{equ:output-model-response}, $y_j(t+1)$ can be further partitioned as
\begin{align}\label{equ:A-output-model-response-ahead-partition}
    y_j (t+1) & = \sum\limits_{i=0}^{t} G_u (t+1,i) U_j (i)  \nonumber \\
                  & + \sum\limits_{i=0}^{t} G_v (t+1,i) v_j(i+1),
\end{align}
where $U_j (t) =[u_j(t), \cdots, u_j(t-n_b+1) ]^T$.

Since $\| G(t,i) \|^2 = \| G_u(t,i) \|^2 + | G_v(t,i) |^2$, then it follows from \eqref{equ:A-bounded-unit-response} that there exist two finite constants $d_{g,u}$ and $d_{g,v}$ such that for each time $t=1, \cdots, N_d$,
\begin{subequations}\label{equ:A-bounded-unit-response-partioned}
\begin{align}
  & \sum\limits_{i=0}^{t-1} \| G_u (t,i) \| \leq d_{g,u}, \\
  & \sum\limits_{i=0}^{t-1} | G_v (t,i) | \leq d_{g,v}, \\
  & d_{g,u}^2 + d_{g,v}^2 = d_{g}^2.
\end{align}
\end{subequations}

Note that the learning controller \eqref{equ:ILC-law-applied} can be rewritten as the following matrix-vector form:
\begin{align}\label{equ:A-ILC-law-matrix-vector}
U_j(t) & = U_{j-1}(t) + \bdiag \left( (\hat{\theta}_{c,j}^{\TR} (t))^T, \cdots, \right. \nonumber \\
           & \qquad \left. (\hat{\theta}_{c,j}^{\TR} (t-n_b+1))^T \right) \bar{E}_{j-1}(t+1),
\end{align}
where $\bdiag(\cdot)$ denotes the block diagonal matrix, and
\begin{align}
  \bar{E}_{j-1}(t+1) = [ E_{j-1}^T(t+1), \cdots, E_{j-1}^T(t-n_b+2) ]^T. \nonumber
\end{align}

Moreover, we let $ \bar{E}_{j-1}(t+1) =\bdiag( \mathcal{E}, \cdots, \mathcal{E} ) \bar{\xi}_{j-1} (t+1)$ with
\begin{subequations}
 \begin{align}
    & \mathcal{E} = \left(
                \begin{array}{ccccc}
                  -1        & 0        & 0         & \cdots & 0 \\
                  -1         & -1       & 0         & \cdots & 0 \\
                  \vdots & \vdots & \vdots & \ddots & \vdots \\
                  -1         & -1       & -1         &  \cdots         & -1 \\
                \end{array}
              \right) \in \mathbb{R}^{n_c \times n_c}, \nonumber \\
   & \bar{\xi}_{j-1}(t+1) = \big[ \xi_{j-1}^T(t+1), \cdots, \xi_{j-1}^T(t-n_b+2) ]^T, \nonumber \\
   & \xi_{j-1}(t+1) = \big[ -e_{j-1}(t+1), \Delta e_{j-1}(t+1), \nonumber \\
   & \quad\quad\quad\quad\quad\quad \cdots, \Delta e_{j-n_c+1}(t+1) \big]^T, \nonumber
\end{align}
\end{subequations}
and $\Delta$ denotes the backward difference operator of a variable in the iteration domain, e.g., $\Delta e_{j-1}(t+1) = e_{j-1}(t+1)- e_{j-2}(t+1)$. Then for each iteration $j=1, \cdots, N_e$ and time $t=1, \cdots, N_d$, according to \eqref{equ:A-output-model-response-ahead-partition} and \eqref{equ:A-ILC-law-matrix-vector}, the tracking error \eqref{equ:error} is further derived by
\begin{align}\label{equ:A-error-model-response-transformed}
 &  e_j(t+1)  = -\sum\limits_{i=0}^{t} G_u (t+1,i)U_{j-1} (i)  \nonumber \\
 &  - \sum\limits_{i=0}^{t} G_u (t+1,i) \bdiag \left( (\hat{\theta}_{c,j}^{\TR} (i))^T, \cdots, \right. \nonumber \\
 & \qquad \left. (\hat{\theta}_{c,j}^{\TR} (i-n_b+1))^T \right) \bdiag( \mathcal{E}, \cdots, \mathcal{E} ) \bar{\xi}_{j-1} (i+1) \nonumber \\
 &  - \sum\limits_{i=0}^{t} G_v (t+1,i) v_j(i+1) + y_d(t+1).
\end{align}

Taking norms on both sides of \eqref{equ:A-error-model-response-transformed}, it can be obtained that
\begin{align}\label{equ:A-tracking-inequality-norm}
  & |e_j(t+1)|  \leq \| \sum\limits_{i=0}^{t} G_u (t+1,i) \|\,  \|U_{j-1} (i) \|  \nonumber \\
                  & + \| \sum\limits_{i=0}^{t} G_u (t+1,i) \|\,  \max \left\{ \| \hat{\theta}_{c,j}^{\TR} (i) \| \, \|  \mathcal{E} \|_F,  \cdots,  \right. \nonumber \\
                   & \quad\quad\quad \left. \| \hat{\theta}_{c,j}^{\TR} (i-n_b+1) \|  \, \|  \mathcal{E} \|_F \right\}\,  \| \bar{\xi}_{j-1} (i+1) \| \nonumber \\
                  & + | \sum\limits_{i=0}^{t} G_v (t+1,i) | \; |v_j(i+1)| + |y_d(t+1)|,
\end{align}
where $\max \{ \cdot \}$ denotes the maximum value of the elements of a set and $\| \cdot \|_F$ denotes the Frobenius norm of a matrix.

Note that for each iteration $j=1, \cdots, N_e$ and time $t=1, \cdots, N_d$, we have
\begin{align}\label{equ:A-tracking-output-relation}
     \Delta e_{j-1}(t+1)  & = \left( y_d(t+1) - y_{j-1}(t+1) \right) \nonumber \\
                                        & - \left( y_d(t+1) - y_{j-2}(t+1) \right) \nonumber \\
                                        & =  - \Delta y_{j-1}(t+1),
\end{align}
and according to \eqref{equ:A-output-model-response-ahead-partition}, taking norms on both sides of \eqref{equ:A-tracking-output-relation}, one has that
\begin{align}\label{equ:A-error-difference-bound}
   & | \Delta e_{j-1}(t+1) | = | \Delta y_{j-1}(t+1) | \nonumber \\
   & \leq  \| \sum\limits_{i=0}^{t} G_u (t+1,i)\|  \left( \|U_{j-1} (i)\|  + \|U_{j-2} (i)\| \right) \nonumber \\
   &  +| \sum\limits_{i=0}^{t} G_v (t+1,i) |  \left( | v_{j-1}(i+1) |  + | v_{j-2}(i+1) | \right).
\end{align}
Moreover, according to Assumption \ref{asmp:bounded-controller},  for each iteration $j=1, \cdots, N_e$ and time $t=1, \cdots, N_d$, we have that
\begin{subequations}\label{equ:A-bounded-uvtheta}
\begin{align}
   & \sup\limits_{ 1 \leq t \leq N_d} \|U_{j}(t) \| = \sqrt{n_b} d_u, \\
   & \sup\limits_{ 1 \leq t \leq N_d} | v_{j}(t) | = d_v, \\
   & \sup\limits_{ 1 \leq t \leq N_d} \| \hat{ \theta}_{c,j}^{\TR}(t) \| = d_c.
\end{align}
\end{subequations}

Then using the results \eqref{equ:A-bounded-unit-response-partioned} and \eqref{equ:A-bounded-uvtheta}, \eqref{equ:A-error-difference-bound} can be obtained by
\begin{align}\label{equ:A-error-difference-bound-completed}
    | \Delta e_{j-1}(t+1) | \leq 2 \sqrt{n_b} d_{g,u} d_u + 2 d_{g,v} d_v,
\end{align}
which further yields that
\begin{align}\label{equ:A-xi-norm}
   \| \xi_{j-1}(t+1) \| & \leq | e_{j-1}(t+1) | +2 \sqrt{n_c-1} \nonumber \\
                                  & \times \left( \sqrt{n_b} d_{g,u} d_u + d_{g,v} d_v \right).
\end{align}
In addition, with \eqref{equ:A-bounded-uvtheta}, we can get
\begin{align}\label{equ:A-block-norm}
   & \max \left\{ \| \hat{\theta}_{c,j}^{\TR} (i) \| \, \|  \mathcal{E} \|_F,  \cdots,  \| \hat{\theta}_{c,j}^{\TR} (i-n_b+1) \|  \, \|  \mathcal{E} \|_F \right\} \nonumber \\
   & \leq d_c \|  \mathcal{E} \|_F = d_c \sqrt{\frac{ n_c (n_c+1)}{2}}.
\end{align}
Then, according to Assumption \ref{asmp:bounded-reference} and the obtained results \eqref{equ:A-bounded-unit-response-partioned}, \eqref{equ:A-bounded-uvtheta}, \eqref{equ:A-xi-norm} and \eqref{equ:A-block-norm}, replacing $ \|U_{j-1} (i) \|$, $\| \bar{\xi}_{j-1} (i+1) \|$, $| v_j(i+1) |$ and $| y_d(t+1) |$ in (\ref{equ:A-tracking-inequality-norm}) by their supremums, we can obtain that
\begin{align}\label{}
      &  | e_j(t+1) |  \leq  d_{g,u} d_c \sqrt{\frac{n_b n_c (n_c+1)}{2}} \Big( | e_{j-1}(t+1) |\nonumber \\
      & + 2 \sqrt{n_c-1} \left( \sqrt{n_b} d_{g,u} d_u + d_{g,v} d_v \right) \Big) + \sqrt{n_b}d_{g,u} d_u \nonumber \\
      & + d_{g,v} d_v +d_r,
\end{align}
which further gives that
\begin{align}\label{equ:A-tracking-inequality-iterations}
 & | e_j(t+1) |  \nonumber \\
 & \leq d_{g,u} d_c \sqrt{\frac{n_b n_c (n_c+1)}{2}} | e_{j-1}(t+1) | +d_r \nonumber \\
  &  + \left( 2 d_{g,u} d_{c}\sqrt{\frac{n_b n_c (n_c^2-1)}{2}} +1\right)  \left(\sqrt{n_b} d_{g,u}d_u + d_{g,v} d_v \right) \nonumber \\
  &  \leq \left(  d_{g,u} d_c \sqrt{\frac{n_b n_c (n_c+1)}{2}} \right)^{j-1} | e_1(t+1) | +  \bigg( \!d_r \nonumber \\
  &  +\! \left(\!2 d_{g,u} d_{c} \sqrt{\frac{n_b n_c (n_c^2\!-\!1)}{2}}+1\! \right)  \left(\sqrt{n_b} d_{g,u} d_u + d_{g,v} d_v \right) \!\bigg) \nonumber \\
  &  \times \left( 1- \left( d_{g,u} d_c \sqrt{\frac{n_b n_c (n_c+1)}{2}}  \right)^{j-1}  \right) \nonumber \\
  & \; \Bigg/ \left(1- d_{g,u} d_c \sqrt{\frac{n_b n_c (n_c+1)}{2}} \right) .
\end{align}
Taking limits for \eqref{equ:A-tracking-inequality-iterations} with $j \rightarrow \infty$, and moreover, using the condition (\ref{equ:condition-theorem1}) and Assumption \ref{asmp:bounded-tracking}, we finally get \eqref{equ:tracking-inequality-final}.

\bibliographystyle{unsrt}
\bibliography{database}

	
	
	
	
	
	%
	%
	

\newpage

\begin{wrapfigure}{l}{25mm}
	\includegraphics [width=25mm, height=35mm]{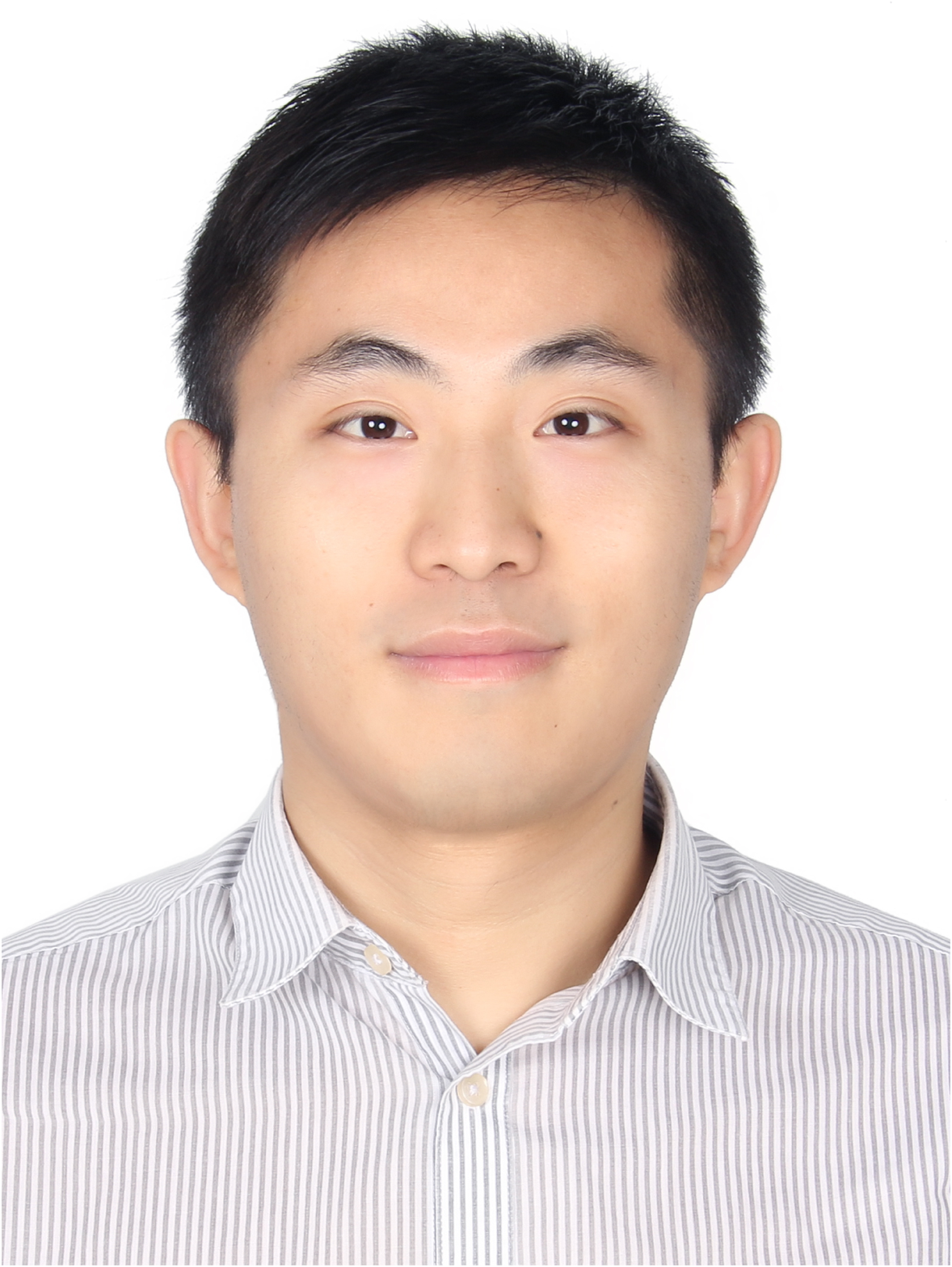}
\end{wrapfigure}
	\textbf{Xian Yu} received his Ph.D. degree in Control Science and Engineering from Beijing Jiaotong University, Beijing, China, in December 2020. He was a visiting researcher in the KIOS Research and Innovation Center of Excellence, University of Cyprus, Nicosia, Cyprus (March 2019 - March 2020).  He is currently working in The Chinese University of Hong Kong, Shenzhen, Guangdong, China, as a Postdoctor (2021 - present).\\
His research interests focus on data-driven control, iterative learning control, system identification, nonlinear systems, multi-agent systems, and their applications. He has served as a reviewer for some journals, such as Automatica, IEEE Transactions on Automatic Control, IEEE Transactions on Systems, Man, Cybernetics: Systems, IEEE Transactions on Industrial Electronics, and IEEE Transactions on Industrial Informatics.

\vspace{10mm}

\begin{wrapfigure}{l}{25mm}
	\vspace{5mm}
	\includegraphics [width=25mm, height=32mm]{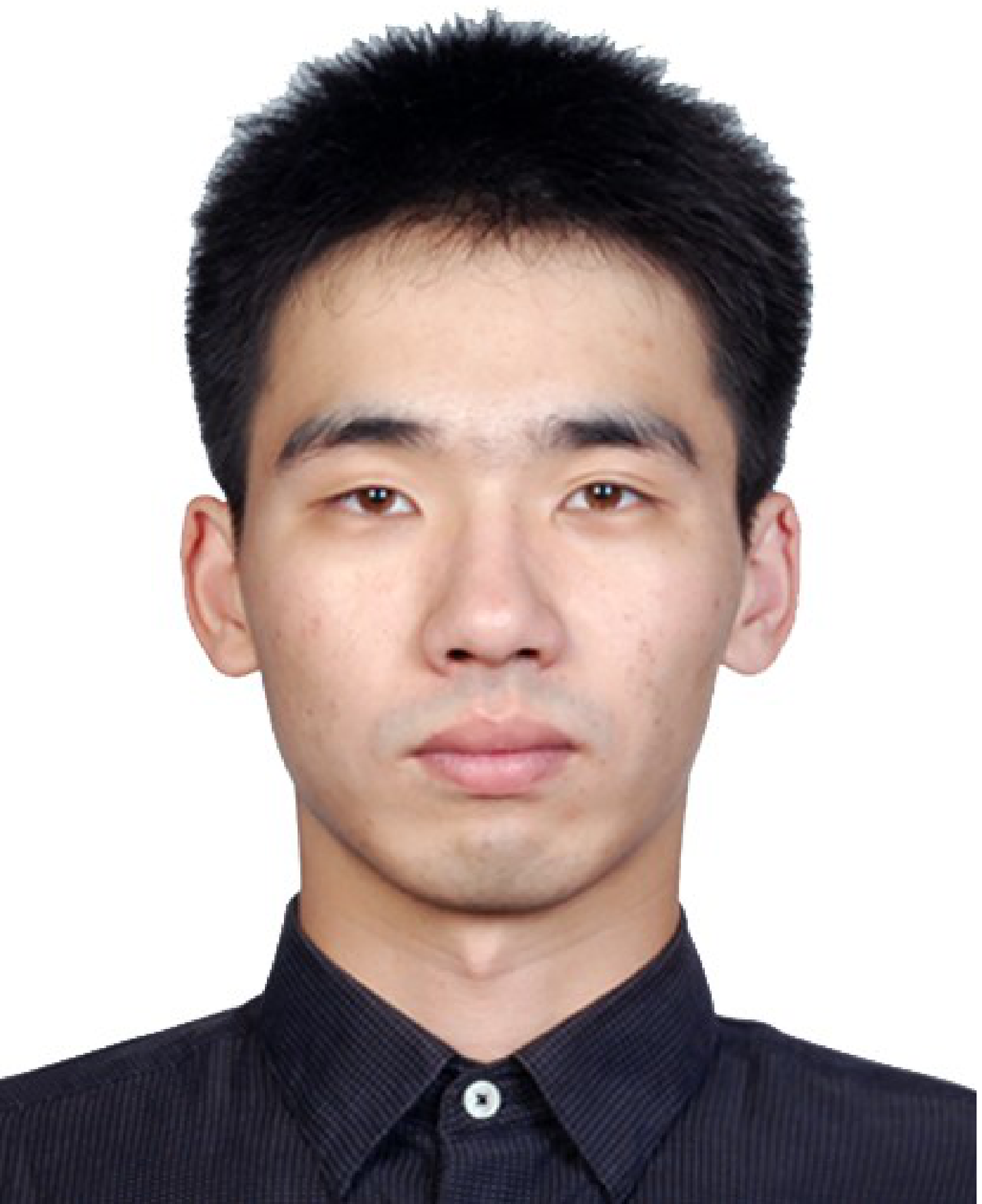}
\end{wrapfigure}
\textbf{Xiaozhu Fang} received his B.S. degree in Mechanical Engineering from Shanghai Jiao Tong University, and his dual M.S. degrees in Mechanical Engineering/Electrical Engineering and Computer Science from University of Michigan, Ann Arbor. He is currently a Ph.D. student in the School of Science and Engineering, The Chinese University of Hong Kong, Shenzhen. His research interests focus on system identification and control theory.

\vspace{10mm}

\begin{wrapfigure}{l}{25mm}
	\vspace{5mm}
	\includegraphics [width=25mm, height=32mm]{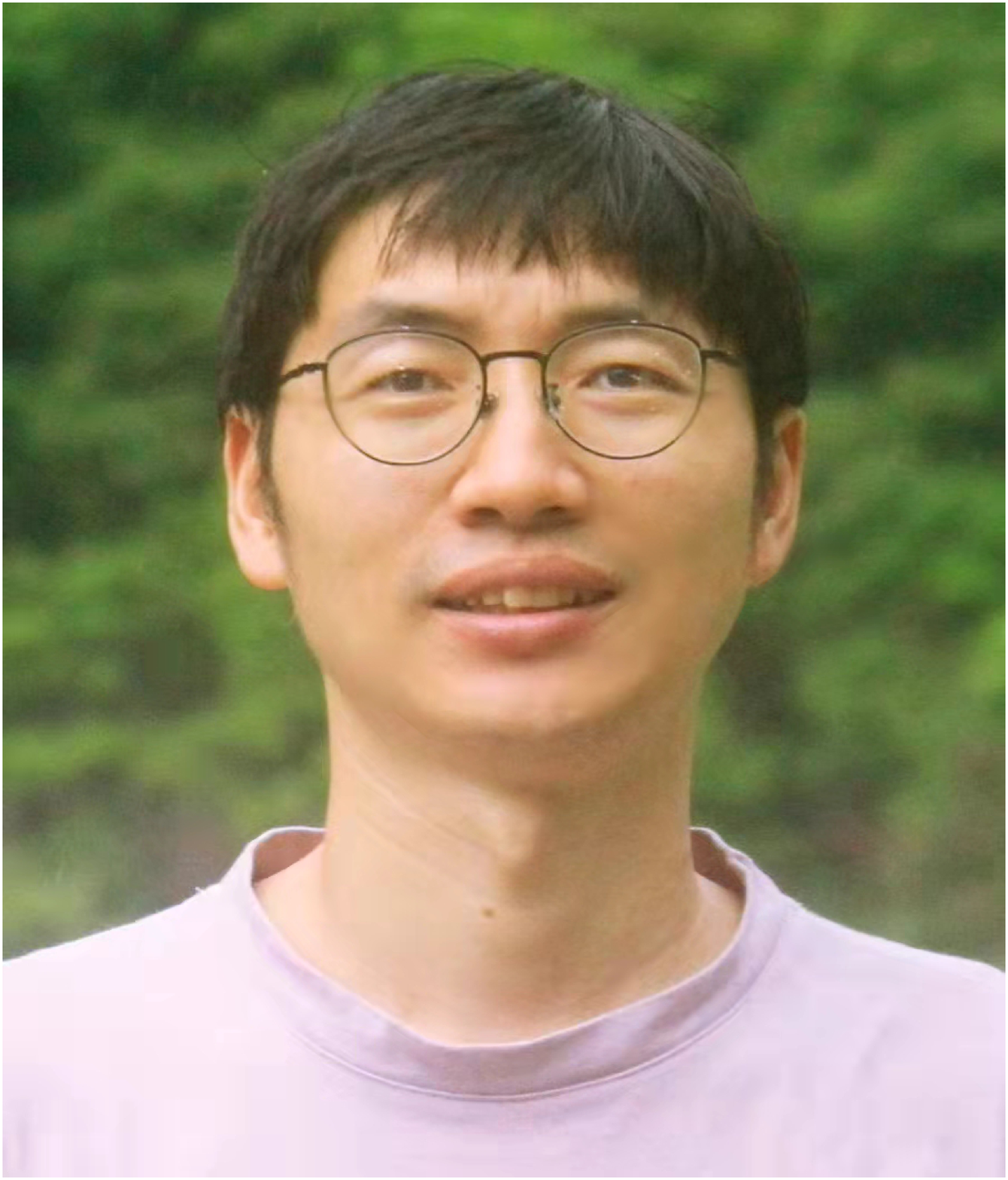}
\end{wrapfigure}
\textbf{Biqiang Mu} received the Bachelor of Engineering degree from Sichuan University and the Ph.D. degree in Operations Research and Cybernetics from the Academy of Mathematics and Systems Science, Chinese Academy of Sciences. He was a postdoc at the Wayne State University, the Western Sydney University, and the Link\"{o}ping University, respectively. He is currently an associate professor at the Academy of Mathematics and Systems Science, Chinese Academy of Sciences. His research interests include system identification, machine learning, and their applications.

\vspace{10mm}

	\begin{wrapfigure}{l}{25mm}
		\includegraphics [width=25mm, height=30mm]{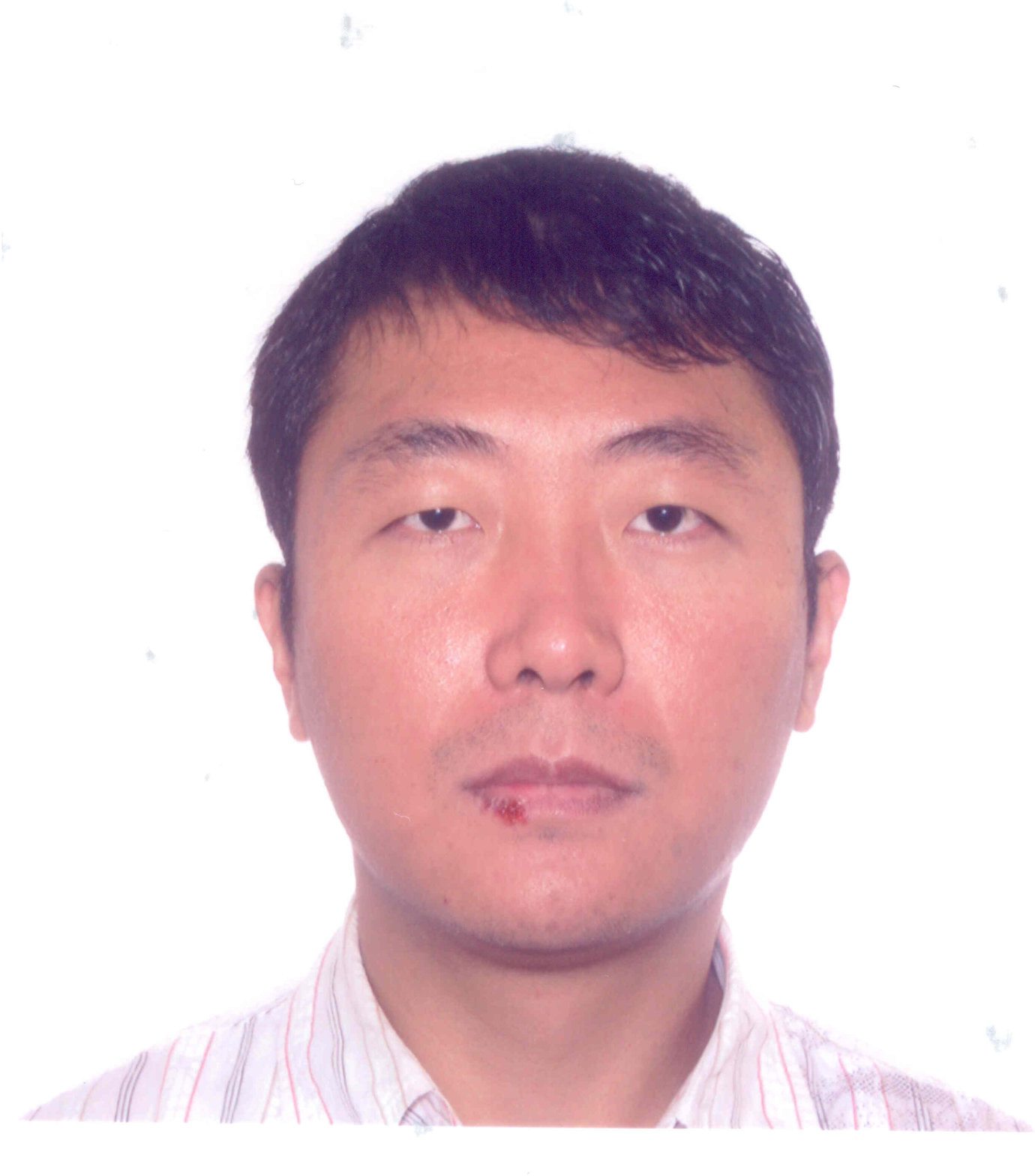}
	\end{wrapfigure}
	\textbf{Tianshi Chen} received his Ph.D. in Automation and Computer-Aided Engineering from The Chinese University of Hong Kong in December 2008. From April 2009 to December 2015, he was working in the Division of Automatic Control, Department of Electrical Engineering, Link\"{o}ping University, Link\"{o}ping, Sweden, first as a Postdoc (April 2009 - March 2011) and then as an Assistant Professor (April 2011 - December 2015). In May 2015, he received the Oversea High-Level Young Talents Award of China, and in December 2015, he returned to China and joined the Chinese University of Hong Kong, Shenzhen, as an Associate Professor.\\
He has been mainly working in the area of systems and control with focus on system identification, state inference, automatic control, and their applications. He is an associate editor for Automatica (2017-present), and also served as an associate editor for System $\&$ Control Letters (2017-2020), and IEEE Control System Society Conference Editorial Board (2016-2019). He was a plenary speaker at the 19th IFAC Symposium on System Identification, Padova, Italy, 2021, and he is a coauthor of the book ``Regularized System Identification - Learning Dynamic Models from Data''.

\end{document}